\documentclass[11pt]{article} 
\usepackage{graphicx,amssymb,mathrsfs,array,multirow}  
\begin{document}  
%\pagestyle{empty}  
%\begin{flushright}  
%\texttt{hep-ph/yymmnnn}\\  
%\end{flushright}  
 
\vskip 30pt

%opening
\begin{center}  
{\Large{\bf 
Scotogenic generation of realistic neutrino mixing with D5
}}\\
\vspace*{1cm}  
\renewcommand{\thefootnote}{\fnsymbol{footnote}}  
{ {\sf Soumita Pramanick$^{1}$\footnote{email: soumita509@gmail.com}}
} \\  
\vspace{10pt}  
{\small  {\em $^1$
Physical Research Laboratory, Ahmedabad, Gujarat, 380009, India
\\
}}
\normalsize  

\normalsize

\end{center}  

%%%%%%%%%%%%%%%%%%%%%%%%%%%%%%%%%%%% Abstract %%%%%%%%%%%%%%%%%%%%%%%%%%%%%%%%%%%%%%%%%%
\begin{abstract} 
\textit{
A mechanism of radiative generation of realistic neutrino mixing at one-loop level with
$D5\times Z_2$ is presented in this paper. The process is demonstrated in two set-ups
using $D5\times Z_2$ symmetry viz. Model 1 and Model 2. Two right-handed neutrinos are 
present in both the models. In both Model 1 and Model 2, when mixing between these two right-handed neutrinos are maximal,
one can produce the form of the left-handed Majorana neutrino mass 
matrix corresponding to $\theta_{13}=0$, $\theta_{23}=\pi/4$
and any value of $\theta_{12}^0$ associated with 
Tribimaximal (TBM), Bimaximal (BM), Golden Ratio (GR) or other mixings.
Small shift from maximal mixing between the two right-handed neutrino states can generate 
non-zero $\theta_{13}$, deviation of $\theta_{23}$ from $\pi/4$ and corrections 
to the solar mixing $\theta_{12}$ in one step for both Model 1 and Model 2.
In both the models, two $Z_2$ odd inert $SU(2)_L$ doublet scalars are present.
The lightest between these two scalars can be a viable dark matter candidate for both Model 1 and Model 2. 
}

\end{abstract}  

\renewcommand{\thesection}{\Roman{section}} 
\setcounter{footnote}{0} 
\renewcommand{\thefootnote}{\arabic{footnote}} 
\noindent

%%%%%%%%%%%%%%%%%%%%%%%%%%%%%%%%%%%% Introduction %%%%%%%%%%%%%%%%%%%%%%%%%%%%%%%%%%%%%%%%%%

\section{Introduction}
\label{Introduction_d5}
Neutrinos oscillate as they are massive. The Pontecorvo, Maki, Nakagawa, Sakata --
PMNS -- matrix connecting the neutrino mass eigenstates with its flavour eigenstates is given by:
\begin{eqnarray}
U = \left(
          \begin{array}{ccc}
          c_{12}c_{13} & s_{12}c_{13} & -s_{13}e^{-i\delta}  \\
 -c_{23}s_{12} + s_{23}s_{13}c_{12}e^{i\delta} & c_{23}c_{12} +
s_{23}s_{13}s_{12}e^{i\delta}&  -s_{23}c_{13}\\
 - s_{23}s_{12} - c_{23}s_{13}c_{12}e^{i\delta}&  s_{23}c_{12} -
c_{23}s_{13}s_{12}e^{i\delta} & c_{23}c_{13} \end{array} \right)
\;\;.
%\nonumber
\label{U_PMNS}
\end{eqnarray}
The $c_{ij}$ and $s_{ij}$ in Eq. (\ref{U_PMNS}) denote $\cos \theta_{ij}$ and $\sin \theta_{ij}$
respectively. It is noteworthy that the neutrino mass eigenstates are non-degenerate.

The observation of non-zero $\theta_{13}$ by short-baseline reactor anti-neutrino experiments \cite{t13_d5}
in 2012 led to important consequences. It must also be noted that although $\theta_{13}$ is non-zero,
the value of $\theta_{13}$ is small compared to the other two mixing angles. Before the discovery of non-zero $\theta_{13}$,
models with leptonic mixings characterized by $\theta_{13}=0$, $\theta_{23}=\pi/4$ 
such as Tribimaximal (TBM), Bimaximal (BM), Golden Ratio (GR) mixings (now onwards in this paper we will refer to these
mixings collectively as popular lepton mixings) were extensively studied. In order to construct TBM, BM and GR mixing patterns one has 
to vary $\theta_{12}^0$
to the particular values mentioned in Table (\ref{tab1}) keeping $\theta_{13}=0$ and $\theta_{23}=\pi/4$ fixed.
Thus, a common structure for all the popular lepton mixings can be easily obtained by putting $\theta_{13}=0$ and $\theta_{23}=\pi/4$ in Eq. (\ref{U_PMNS}):
\begin{equation}
U^0=
\pmatrix{\cos \theta_{12}^0 & \sin \theta_{12}^0  & 0 \cr -\frac{\sin
\theta_{12}^0}{\sqrt{2}} & \frac{\cos \theta_{12}^0}{\sqrt{2}} &
-{1\over\sqrt{2}} \cr
-\frac{\sin \theta_{12}^0}{\sqrt{2}} & \frac{\cos
\theta_{12}^0}{\sqrt{2}}  & {1\over\sqrt{2}}},
\label{mix0_d5}
\end{equation}
with $\theta_{12}^0$ for TBM, BM and GR displayed in Table (\ref{tab1}).

The 3$\sigma$ global fit \cite{nufit2022_d5, Gonzalez_d5, Valle_d5} of the three neutrino mixing angles viz. $\theta_{12}$, $\theta_{13}$ and $\theta_{23}$ as per NuFIT5.2 of 2022 \cite{nufit2022_d5}:
\begin{eqnarray}
\theta_{12} &=&(31.31 - 35.74)^\circ, \nonumber \\
\theta_{23} &=& (39.6-51.9)^\circ \,, \nonumber \\
\theta_{13} &=& (8.19-8.89)^\circ.
\label{mix3sigma}
\end{eqnarray}

%--------------------------- 
\begin{table}[tb]
\begin{center}
\begin{tabular}{|c|c|c|c|}
\hline
  Model &TBM &BM & GR \\ \hline
$\theta^0_{12}$ & 35.3$^\circ$ & 45.0$^\circ$  & 31.7$^\circ$ \\ \hline
\end{tabular}
\end{center}
\caption{\sf{$\theta_{12}^0$ values of various poular lepton mixings viz. TBM, BM and GR mixings.}}
\label{tab1}
\end{table}
%---------------------
Therefore popular lepton mixings are in sharp disagreement with the non-zero $\theta_{13}$.
Extensive model-building activities have been taking place after the observation of non-zero $\theta_{13}$
so as to incorporate it in the popular lepton mixing frameworks. In \cite{brp_d5}, attempts have been made to trace a common
origin for the two small quantities viz. non-zero $\theta_{13}$ and the solar mass splitting ($\Delta m^2_{solar}$).
In \cite{pr_d5}, a two-component neutrino mass matrix scenario was considered where the dominant one was associated
with oscillation parameters such as the atmospheric mass splitting ($\Delta m^2_{atm}$) and $\theta_{23}=\pi/4$
whereas the other oscillation parameters such as the non-zero $\theta_{13}$, $\theta_{12}$, $\Delta m^2_{solar}$ as well as deviation of $\theta_{23}$ from $\pi/4$ could be generated perturbatively with the help of smaller see-saw \cite{seesaw_d5}
contribution\footnote{Similar earlier works can be found in \cite{old_d5}.}. 
In some studies \cite{othermodels_d5, LuhnKing_d5}, vanishing $\theta_{13}$ was obtained
with the help of different symmetries and one could yield non-zero $\theta_{13}$ by adding some perturbations
to the symmetric structures.

The popular lepton mixings were corrected by devising a two-component Lagrangian mechanism at tree-level with the help of 
discrete flavour symmetries such as $A4$, $S3$ in \cite{ourS3_d5, newA4_d5}. The type II see-saw dominant contribution
of the Lagrangian corresponding to the popular lepton mixings was corrected by a smaller type I see-saw subdominant contribution
in these models. An exactly similar study specific  to the no solar mixing (NSM) i.e., $\theta_{12}^0=0$ case using 
$A4$ symmetry\footnote{Solar splitting was absent in the dominant type II see-saw component thereby allowing us to use
degenerate perturbation theory for generation of large solar mixing in \cite{ourA4_d5}.} 
can be found in \cite{ourA4_d5}. A study of generating TBM radiatively with $A4$ can be found in \cite{Ma_rad_d5}.
Some recent studies of realistic neutrino mixings are present in \cite{newMa_d5, tanimoto_d5}. For some earlier works on scotogenic models, see \cite{scotogenic_Prof_Valle_d5}. 

In this paper, radiative generation of realistic neutrino mixing with $D5\times Z_2$ symmetry 
will be presented\footnote{The discrete symmetry $D5$ is briefly discussed in Appendix \ref{D5group}.}.
Some earlier works on $D5$ symmetry can be found in \cite{Ma_d5, Lindner_d5}.

The prime intent of this work is to use $D5\times Z_2$ symmetry to radiatively\footnote{A detailed 
analysis of radiative neutrino mass models can be found in \cite{radreview_d5}.} produce:
\begin{enumerate}
\item The form of the mixing matrix of popular lepton mixing kind given in Eq. (\ref{mix0_d5}) corresponding to
$\theta_{13}=0$, $\theta_{23}=\pi/4$ and $\theta_{12}^0$ of any of the options shown in Table (\ref{tab1}).
\item Realistic neutrino mixings i.e., non-zero $\theta_{13}$, deviation of $\theta_{23}$ from $\pi/4$ and 
minute corrections to solar mixing $\theta_{12}$.
\end{enumerate}

Here the neutrino masses and mixings are generated at one-loop level using $D5\times Z_2$ symmetry.
We try to demonstrate the mechanism in two model set-ups viz. Model 1 and Model 2. 
The two models vary in the $D5$ quantum number assignments of the fields. Needless to mention, 
both the models contain same types of fields which only differ in their $D5$ charges and both the models 
follow the same modus operandi as discussed below:

In both the models, two right-handed neutrinos are present and when the mixing between these two right-handed
neutrino states are maximal (i.e., $\pi/4$), the structure of the leptonic mixing matrix corresponding to the popular
lepton mixings characterized by $\theta_{13}=0$ and $\theta_{23}=\pi/4$ as displayed in Eq. (\ref{mix0_d5}) can be generated.
We will see in course of our discussion that in both the models, a tiny deviation from maximal 
mixing in between the two right-handed neutrinos
can successfully yield the realistic mixings i.e., non-zero $\theta_{13}$, shifts of the atmospheric mixing angle $\theta_{23}$ from $\pi/4$ and slight changes in solar mixing $\theta_{12}$, in a single stroke. The process necessitates two $Z_2$
odd inert $SU(2)_L$ doublet scalars $\eta_i$ ($i=1,2$), the lightest among which can serve as a dark matter candidate.
A similar method was implemented in a scotogenic $S3$ symmetric model \cite{radS3_d5}.
In the scotogenic $A4$ model presented in \cite{radA4_d5}, small mass splitting between two-right handed neutrino states were
used instead of tweaking the maximal mixing between the right-handed neutrinos to generate realistic mixings such as non-zero 
$\theta_{13}$, deviation of $\theta_{23}$ from $\pi/4$ and 
minute modifications of $\theta_{12}$.

In Section \ref{firstmodel} we discuss Model 1 in details, followed by a comprehensive analysis of Model 2 in Section \ref{secondmodel} and finally we
conclude with a brief account of our findings in Section \ref{conclusions}. In Appendix \ref{D5group} we discuss about the discrete group $D5$
in general and in Appendix \ref{scalarpotentials} the scalar potentials of both Model 1 and Model 2 have been scrutinized\footnote{We will see in Appendix \ref{scalarpotentials}, that although the $D5$ quantum numbers of the scalars 
in Model 1 and Model 2 are different eventually the scalar potentials for both Model 1 and Model 2 came out to be the same.}.

\section{The $D5\times Z_2$ Models:}
\label{d5model_in_general}
The left-handed Majorana neutrino mass matrix in mass basis is given by $M^{mass}_{\nu L}=$ $diag\; (m_1, m_2, m_3)$.
The flavour basis form of this $M^{mass}_{\nu L}$ for 
$\theta_{13}=0$, $\theta_{23}=\pi/4$ and $\theta_{12}^0$ specific to any of the popular lepton mixing alternatives displayed in Table (\ref{tab1}) can be easily obtained by using the common structure of leptonic mixing
matrix $U^0$ given in Eq. (\ref{mix0_d5}):
\begin{equation}
M^{flavour}_{\nu L}=U^0 M^{mass}_{\nu L} U^{0T}=\pmatrix{a & c & c \cr
c & b & d \cr c & d & b}.
\label{abc_d5}
\end{equation} 
The $a,b,c$ and $d$ in Eq. (\ref{abc_d5}) are:
\begin{eqnarray}
a&=& m_1\cos^2 \theta_{12}^0+m_2\sin^2 \theta_{12}^0\nonumber\\
b&=&\frac{1}{2}\left(m_1\sin^2 \theta_{12}^0+m_2\cos^2 \theta_{12}^0+m_3\right)\nonumber\\
c&=&\frac{1}{2\sqrt{2}}\sin 2\theta_{12}^0 (m_2-m_1)\nonumber\\
d&=&\frac{1}{2}\left(m_1\sin^2 \theta_{12}^0+m_2\cos^2 \theta_{12}^0-m_3\right).
\label{abcd_d5}
\end{eqnarray}
Therefore, 
\begin{equation}
\tan 2\theta_{12}^0=\frac{2\sqrt 2 c}{b+d-a}.
\end{equation}
For neutrino masses to be non-degenerate and realistic, these quantities $a,b,c$ and $d$ have to be non-zero. 

As discussed above, for both the models the first step will be to obtain the structure of $M^{flavour}_{\nu L}$
in Eq. (\ref{abc_d5}) radiatively at one-loop level by appropriately choosing the 
$D5\times Z_2$ quantum numbers of the fields in the models. As already mentioned, this will require maximal mixing between the 
two right-handed neutrino states.
Once the form $M^{flavour}_{\nu L}$ in Eq. (\ref{abc_d5}) is achieved, one has to tweak 
the maximal mixing between these two right-handed neutrinos
to get the realistic neutrino mixings.
The same procedure will be followed for both Model 1 and Model 2 which differ from each other in terms of $D5$ quantum number
assignments of the particles present in the models. Let us now analyze the models one by one.

\subsection{Model 1}
\label{firstmodel}
\begin{figure}[tbh]
\begin{center}
\includegraphics[scale=0.22,angle=0]{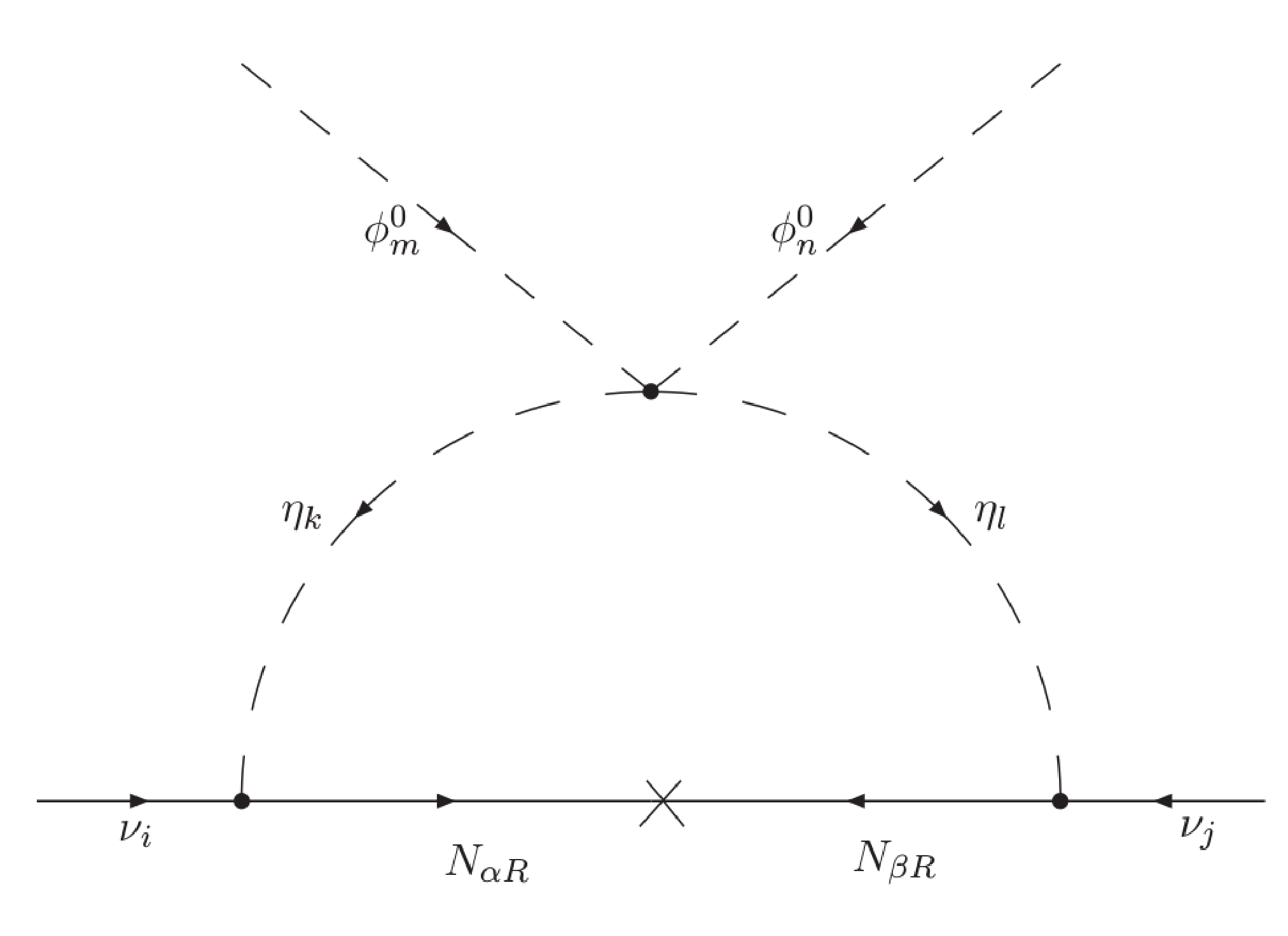}
\end{center}
\caption{\sf  Scotogenic neutrino mass generation at one-loop level with $D5 \times Z_2$ symmetry.
This figure is responsible for neutrino mass generation in both Model 1 and Model 2.
}
\label{radfig_d5} 
\end{figure}

In Model 1, the three left-handed lepton $SU(2)_L$ doublets $L_{\zeta_L}\equiv(\nu_\zeta \ \ \zeta^-)_L^T$
where $\zeta=e,\mu,\tau$ are present. Two of these left-handed lepton $SU(2)_L$ doublets i.e., $L_{\mu_L}$ and $L_{\tau_L}$ 
transform as $2_1$ under $D_5$ and the $D_5$ charge of $L_{e L}$ is $1_1$. In addition to these, two Standard Model (SM) gauge singlet right-handed neutrinos $N_{\alpha R}$,
($\alpha=1,2$) are also present in this model that transform as $2_2$ under $D5$. 
The scalar sector comprises of two inert $SU(2)_L$ doublet scalars, 
$\eta_i\equiv(\eta_i^+, \eta_i^0)^T$, $(i=1,2)$ transforming as $2_2$ of $D5$ denoted by $\eta$.
The other two $SU(2)_L$ doublet scalars i.e., $\Phi_j\equiv(\phi_j^+, \phi_j^0)^T$, $(j=1,2)$
having $D5$ charge $2_2$ is designated by $\Phi$.
Apart from $D5$, the unbroken $Z_2$ present in the model plays a crucial role under which 
all the fields except the right-handed neutrinos and scalar $\eta$ are even.  The $\phi_j$
being $Z_2$ even can get vacuum expectation value (vev) after spontaneous symmetry breaking (SSB)
whereas $\eta_i$ being odd under $Z_2$ cannot get vev. Let $v_j$ denote the vevs of $\phi_j^0$ i.e., $\langle \Phi_j\rangle\equiv v_j$, $(j=1,2)$. The particle content of Model 1 along with their specific quantum numbers are displayed
in Table (\ref{particles_model1}). In this model, we are dealing with the neutrino sector only. We are working in a basis in which the charged lepton mass matrix is diagonal and the entire mixing originates from the neutrino sector.
%--------------------------- 
\begin{table}[tb]
\begin{center}
\begin{tabular}{|c|c|c|c|}
\hline
\sf{Leptons} & $SU(2)_L$ & $D5$ & $Z_2$ \\ \hline
 & & &  \\ 
$L_{e_L}\equiv\pmatrix{\nu_e& e^-}_L$ & $2$ & $1_1$ & $1$ \\
 & & &  \\ 
 \hline
 & & &  \\ 
$L_{\zeta_L}\equiv\pmatrix{
\nu_\mu & \mu^- \cr \nu_\tau & \tau^- }_L$& $2$ & $2_1$ & $1$ \\
 & & &  \\ 
 \hline
 & & &  \\ 
$N_{\alpha R}\equiv \pmatrix{N_{1R}\cr
N_{2R}} $ & $1$ & $2_2$ & $-1$ \\ 
 & & &  \\ 
\hline
\hline
\sf{Scalars}& $SU(2)_L$ & $D5$ & $Z_2$ \\ \hline
 & & &  \\ 
$\Phi \equiv \pmatrix{\phi_1^+ & \phi_1^0\cr
\phi_2^+ & \phi_2^0 }$ & $2$ & $2_2$ & $1$ \\ 
 & & &  \\ 
\hline
 & & &  \\ 
$\eta\equiv \pmatrix{\eta_1^+ & \eta_1^0\cr
\eta_2^+ & \eta_2^0 }$ & $2$ & $2_2$ & $-1$ \\ 
 & & &  \\ 
\hline
\end{tabular}
\end{center}
\caption{\sf{Fields in Model 1 with their corresponding quantum numbers. Here we restrict ourselves to the
neutrino sector only.}}
\label{particles_model1}
\end{table}
%---------------------
\vskip 2pt
At one-loop level, neutrino mass can be obtained radiatively\footnote{Later on in Section (\ref{secondmodel}) we will find that the same diagram in Fig. (\ref{radfig_d5}) can generate neutrino mass radiatively at one-loop for Model 2 as well. Thus Fig. (\ref{radfig_d5}) serves the purpose of neutrino mass generation radiatively at one-loop for both Model 1 and Model 2. } from Fig. (\ref{radfig_d5}) in Model 1.
The relevant part of the $D5\times Z_2$ conserving scalar potential\footnote{At the four-point scalar vertex in Fig. (\ref{radfig_d5}), two $\eta$ are created and two $\phi$ are annihilated. Thus only the terms of the scalar potential
of $(\eta^\dagger\phi)(\eta^\dagger\phi)$ kind contribute to the left-handed Majorana neutrino mass matrix.
The complete scalar potential
for both Model 1 and Model 2 are same and can be found in Appendix \ref{scalarpotentials}.} that contributes to the 
left-handed Majorana neutrino mass matrix via the four-point scalar vertex in Fig. (\ref{radfig_d5}):
\begin{eqnarray}
V_{relevant}&\supset& \lambda_1 \left[ \left\{(\eta_2^\dagger \phi_2+\eta_1^\dagger \phi_1)^2 \right  \}  + h.c.\right]+ \lambda_2\left[ \left\{(\eta_2^\dagger \phi_2- \eta_1^\dagger \phi_1 )^2 \right \} +h.c.
\right]\nonumber\\
&+&\lambda_3\left[ \left\{(\eta_1^\dagger \phi_2)(\eta_2^\dagger \phi_1)+(\eta_2^\dagger \phi_1)(\eta_1^\dagger \phi_2)\right  \} 
+h.c.\right].
\label{potential_d5}
\end{eqnarray}
We take all the quartic couplings $\lambda_j$ ($j=1,2,3$) to be real.
%---------------------------------------------------------------------------
%---------------------------------------------------------------------------

All symmetries under consideration are conserved at all the three vertices of Fig. (\ref{radfig_d5}).
The $D5\times Z_2$ conserving Dirac vertices are given by:
\begin{equation}
{\mathscr{L}_{Yukawa}}|_{\rm{(Model\,1)}}= y_1\left[(\overline{N}_{2R}\eta_2^0+\overline{N}_{1R}\eta_1^0)\nu_e\right]
+y_2\left[(\overline{N}_{1R}\eta_2^0)\nu_\tau+ (\overline{N}_{2R}\eta_1^0)\nu_\mu\right]+h.c.
\label{yukawa_d5_model1}
\end{equation}
It is imperative to note that owing to the difference in the $D_5$ quantum numbers of the left-handed neutrinos
as displayed in Table (\ref{particles_model1}), the Yukawa couplings of $\nu_\zeta$ for ($\zeta=\mu,\,\tau$) transforming as $2_1$
of $D5$ is different from that of $\nu_e$ with $D5$ quantum number $1_1$ in Eq. (\ref{yukawa_d5_model1}). Precisely, the Yukawa couplings of $\nu_\zeta$ for ($\zeta=\mu,\,\tau$) and $\nu_e$ are denoted by $y_2$ and $y_1$ respectively.

In the right-handed neutrino sector we have two right-handed neutrinos, $N_{1R}$ and $N_{2R}$ transforming as $2_2$
under $D_5$. The $D5 \times Z_2$ invariant direct mass term for the right-handed neutrinos:
\begin{equation}
\mathscr{L}_{right-handed \, neutrinos}= \frac{1}{2} m_{R_{12}}
\left[N_{1R}^T C^{-1}N_{2R}+N_{2R}^T C^{-1}N_{1R}\right].
\label{rhnu_model1_d5intact}
\end{equation}
Therefore, in a $D5$ conserving scenario only non-zero off-diagonal entries are allowed in the right-handed Majorana
neutrino mass matrix. To get non-zero diagonal entries, one can add soft $D5$ breaking terms:
\begin{equation}
\mathscr{L}_{soft}= \frac{1}{2}\left[ m_{R_{11}}
N_{1R}^T C^{-1}N_{1R}+m_{R_{22}}N_{2R}^T C^{-1}N_{2R}\right]
\label{rhnu_model1_d5softbreak}
\end{equation}
This enables us to write the right-handed neutrino mass matrix as the following:
\begin{equation}
M_{\nu_R}=\frac{1}{2}\pmatrix{m_{R_{11}} & m_{R_{12}}\cr
m_{R_{12}} & m_{R_{22}}
}.
\label{mrh_d5}
\end{equation}
This symmetric form of the right-handed neutrino mass matrix\footnote{We will see later on in Section (\ref{secondmodel}), that although the $D5$ charges of the right-handed neutrinos in Model 1 differ from that in Model 2, one can still arrive to this form of the right-handed Majorana neutrino mass matrix shown in Eq. (\ref{mrh_d5}) in Model 2 as well.} is indicative of its Majorana nature.

At this stage it is prudent to have a look at the dark matter candidates present in the model.
It is a general practice to appoint $Z_2$ symmetry to stabilize the dark matter candidates in the model
and since this model also has a $Z_2$ symmetry, one can immediately perceive the existence of dark matter candidates
in the model. From Table (\ref{particles_model1}), we can see that both the right-handed neutrinos $N_{\alpha R}$ with ($\alpha= 1, \, 2$)
as well as the scalars $\eta$ are odd under $Z_2$. Here, we choose $\eta$ to be lighter than the right-handed neutrinos $N_{\alpha R}$, ($\alpha= 1, \, 2$). 
From $m^2_\eta$ term in Eq. (\ref{totalpotential_d5}) it might seem that the $\eta_i$, ($i=1,2$) scalars are degenerate in mass, but since
$D5$ is already broken softly at the right-handed neutrino mass scale, one can have tiny mass splitting between the two $\eta_i$, ($i=1,2$) scalars. Thus the lightest among the two $\eta_i$, ($i=1,2$) can serve as a dark matter candidate.
 
We now have the essential components of the model ready in our hands to delineate the left-handed Majorana neutrino mass matrix originating from Fig. (\ref{radfig_d5}). The detailed expressions for it will be presented afterwards but for the time being let us have a sketchy idea of how the one-loop diagram \cite{Ma_loop_d5} in Fig. (\ref{radfig_d5}) contributes to the left-handed Majorana mass matrix. At this point, a few simplifying assumptions are required to make the expressions appear less complicated. 
Let some combination of the three quartic couplings $\lambda_1, \, \lambda_2, \lambda_3$ in Eq. (\ref{potential_d5})
be commonly represented by $\lambda$. Furthermore, we neglect the mass splitting between $\eta_1$ and $\eta_2$ and assume $m_0$
to be the common mass of these scalar. Let $\eta_{Rj}$ be the real part of $\eta_j^0$ and $\eta_{Ij}$ be the imaginary part of $\eta_j^0$. We can take the splitting between the masses of $\eta_{Rj}$ and $\eta_{Ij}$ to be proportional to $\lambda v_j$ which can be generally a small quantity. 

Once again, it is crucial to remind ourselves that $\nu_\zeta$ for ($\zeta=\mu,\,\tau$) transform as $2_1$ under $D5$, while
the $D5$ charge of $\nu_e$ is $1_1$. This will play an important role in determining the structure of the left-handed Majorana neutrino mass matrix via the Yukawa couplings given in Eq. (\ref{yukawa_d5_model1}) at the two Dirac vertices in Fig. (\ref{radfig_d5}). Let $m_R$ denote the average mass of the heavy right-handed neutrino states. Hence we can define 
$z\equiv \frac{m_R^2}{m_0^2}$. In order to define $z$, we do not distinguish between masses of the two right-handed neutrinos as the quantity $z$ appears only in the logarithm throughout the analysis. Now we can write the second diagonal entry of 
$M_{\nu L}^{flavour}$ as:
\begin{equation}
{(M_{\nu_L}^{flavour})_{22}}|_{\rm{(Model\,1)}}=\lambda \frac{v_m v_n}{8 \pi^2}
\frac{y_2^2 }{m_{R_{22}}} 
\left[ \ln z -1 \right].
\label{m22_d5_model1}
\end{equation}
This expression in Eq. (\ref{m22_d5_model1}) is valid in limit $m_R^2>>m_0^2$.
From Eq. (\ref{yukawa_d5_model1}), it is clear to us that $\nu_\mu$ couples to $N_{2R}$ only with Yukawa coupling $y_2$.
Therefore for $(M_{\nu_L}^{flavour})_{22}$, $\nu_\mu$ couples to $N_{2R}$ at both the Dirac vertices.
Thus $(M_{\nu_L}^{flavour})_{22}$ receives contributions only from $m_{R_{22}}$ with $y_2$ being the only Yukawa coupling appearing in it. One can obtain the expression for $(M_{\nu_L}^{flavour})_{33}$ in an exactly similar way just by replacing $m_{R_{22}}$ 
by $m_{R_{11}}$ in Eq. (\ref{m22_d5_model1}).

Moving on to the off-diagonal (2,3) entry of $M_{\nu_L}^{flavour}$ viz. $(M_{\nu_L}^{flavour})_{23}$. Here $\nu_\mu$ contributes at one of the Dirac vertices while $\nu_\tau$ contributes at the other one. As we know from Eq. (\ref{yukawa_d5_model1}), $\nu_\mu$ and $\nu_\tau$ couple with $N_{2R}$ and $N_{1R}$ respectively leading to existence of $N_{2R}$ at one of the Dirac vertices and $N_{1R}$ at the other one. Thus $(M_{\nu_L}^{flavour})_{23}$
will receive contributions from off-diagonal entries $m_{R_{12}}$ of the right-handed neutrino mass matrix $M_{\nu_R}$
shown in Eq. (\ref{mrh_d5}) together with diagonal entries $m_{R_{11}}$ and $m_{R_{22}}$. It is noteworthy, that the Yukawa coupling appearing at both the Dirac vertices is $y_2$ as can be seen from Eq. (\ref{yukawa_d5_model1}). Therefore we get:
\begin{equation}
{(M_{\nu_L}^{flavour})_{23}}|_{\rm{(Model\,1)}}=\lambda \frac{v_m v_n}{8 \pi^2}
\frac{y_2^2 m_{R_{12}}}{m_{R_{11}}m_{R_{22}}} 
\left[ \ln z -1 \right].
\label{m23_d5_Model1}
\end{equation} 
We have used mass insertion approximation while obtaining Eq. (\ref{m23_d5_Model1}). The same procedure can be followed to obtain the expressions for the (1,1), (1,2) and (1,3) entries of the left-handed Majorana neutrino mass matrix $M_{\nu_L}^{flavour}$.

In order to make the expressions for the elements of $M_{\nu_L}^{flavour}$ look less cumbersome, we absorb everything else in RHS of Eqs. (\ref{m22_d5_model1}) 
and (\ref{m23_d5_Model1}) except the quartic couplings, Yukawa couplings and the vevs in some loop contributing factotrs
denoted by $r_{\alpha\beta}$ as shown below:
\begin{eqnarray}
r_{11}&\equiv&\frac{1}{8 \pi^2 m_{R_{11}}}\left[ \ln z -1 \right],\nonumber\\
r_{22}&\equiv&\frac{1}{8 \pi^2 m_{R_{22}}}\left[ \ln z -1 \right],\nonumber\\
r_{12}&\equiv&\frac{m_{R_{12}}}{ 8 \pi^2 m_{R_{11}}m_{R_{22}}}\left[ \ln z -1 \right].
\label{rij_d5}
\end{eqnarray}
Using Eqs. (\ref{m22_d5_model1}), (\ref{m23_d5_Model1}), (\ref{rij_d5}) and (\ref{potential_d5}), the left-handed Majorana neutrino mass matrix produced radiatively with Fig. (\ref{radfig_d5}) at one-loop level is given by:
\begin{equation}
{M_{\nu_L}^{flavour}}|_{\rm{(Model\,1)}}=\pmatrix{ \chi_1
& \chi_4 & \chi_5 \cr
\chi_4
& \chi_2 & \chi_6 \cr 
\chi_5
& \chi_6 & \chi_3
}.
\label{mgeneral_d5_model1}
\end{equation}
Here,
\begin{eqnarray}
\chi_1&\equiv&y_1^2\left[
4 r_{12} v_1v_2(\lambda_3+\lambda_1-\lambda_2)
+(r_{11}v_1^2+r_{22}v_2^2)(\lambda_1+\lambda_2)
\right]\nonumber\\
\chi_2&\equiv&y_2^2\left[r_{22}(\lambda_1+\lambda_2)v_1^2\right]\nonumber\\
\chi_3&\equiv&y_2^2\left[r_{11}(\lambda_1+\lambda_2)v_2^2\right]\nonumber\\
\chi_4&\equiv& y_1y_2\left[
r_{12}(\lambda_1+\lambda_2)v_1^2+2r_{22}(\lambda_3+\lambda_1-\lambda_2)v_1v_2
\right]\nonumber\\
\chi_5&\equiv& 
y_1y_2\left[
r_{12}(\lambda_1+\lambda_2)v_2^2+2r_{11}(\lambda_3+\lambda_1-\lambda_2)v_1v_2
\right]
\nonumber\\
\chi_6&\equiv& y_2^2\left[
2 r_{12}(\lambda_3+\lambda_1-\lambda_2)v_1v_2
\right].
\label{chidefinition_d5_model1}
\end{eqnarray}
Needless to mention that $\langle \Phi_j\rangle\equiv v_j$ with ($j=1,2$) in Eq. (\ref{chidefinition_d5_model1}).

As already mentioned in Section (\ref{Introduction_d5}), we will at first try to obtain the form of the left-handed Majorana neutrino mass matrix in Eq. (\ref{abc_d5}) characterized by $\theta_{13}=0$, $\theta_{23}=\pi/4$ and $\theta_{12}^0$ corresponding to any of the popular mixing values given in Table (\ref{tab1}). For this we need $\chi_1\ne\chi_2=\chi_3$
together with $\chi_4=\chi_5$. As we can see from Eq. (\ref{chidefinition_d5_model1}), 
one can get this by setting $v_1=v_2=v$ and $r_{11}=r_{22}=r$. 
In the right-handed neutrino mass matrix in Eq. (\ref{mrh_d5}), the condition $r_{11}=r_{22}=r$ will manifest as:
\begin{equation}
M_{\nu_R}=\frac{1}{2}\pmatrix{m_{R_{11}} & m_{R_{12}}\cr
m_{R_{12}} & m_{R_{11}}
}.
\label{mrh2_d5}
\end{equation}
We have used Eq. (\ref{rij_d5}) to calculate Eq. (\ref{mrh2_d5}). A glance at Eq. (\ref{mrh2_d5}) will immediately reveal that the $r_{11}=r_{22}=r$ criterion basically means maximal mixing between the two right-handed neutrino states $N_{1R}$ and $N_{2R}$
as we discussed earlier in Section (\ref{Introduction_d5}). Summing up, in order to get the structure of $M_{\nu_L}^{flavour}$ as in Eq. (\ref{abc_d5}) in Model 1, one has to make $v_1=v_2=v$ and mixing between $N_{1R}$ and $N_{2R}$ maximal i.e., 
$r_{11}=r_{22}=r$. Applying these conditions in the general form of $M_{\nu_L}^{flavour}$ for Model 1 given in Eq. (\ref{mgeneral_d5_model1}) one can get:
\begin{equation}
{M_{\nu_L}^{flavour}}|_{\rm{(Model\,1)}}=v^2\pmatrix{ y_1^2[4r_{12}\lambda_{123} +2r\lambda_{12}]
& y_1y_2[r_{12}\lambda_{12} +2r\lambda_{123}]
& y_1y_2[r_{12}\lambda_{12} +2r\lambda_{123}]\cr
y_1y_2[r_{12}\lambda_{12} +2r\lambda_{123}]
& y_2^2r\lambda_{12}& y_2^2(2r_{12}\lambda_{123})\cr 
y_1y_2[r_{12}\lambda_{12} +2r\lambda_{123}]
& y_2^2(2r_{12}\lambda_{123}) & y_2^2r\lambda_{12}
}.
\label{mchoicemaxmix_d5_model1}
\end{equation} 
In Eq. (\ref{mchoicemaxmix_d5_model1}) we have introduced the compact notation: $\lambda_{12}\equiv\lambda_1+\lambda_2$ and $\lambda_{123}\equiv\lambda_3+\lambda_1-\lambda_2$. To achieve the form of $M_{\nu_L}^{flavour}$ mentioned in Eq. (\ref{abc_d5})
from Eq. (\ref{mchoicemaxmix_d5_model1}) within the framework of Model 1 one has to do the following mappings:
\begin{eqnarray}
a&\equiv&y_1^2v^2[4r_{12}\lambda_{123} +2r\lambda_{12}]
= y_1^2v^2[4r_{12}(\lambda_3+\lambda_1-\lambda_2) +2r(\lambda_1+\lambda_2)]\nonumber\\
b&\equiv& y_2^2v^2r\lambda_{12}=y_2^2v^2r(\lambda_1+\lambda_2)\nonumber\\
c&\equiv&y_1y_2v^2[r_{12}\lambda_{12} +2r\lambda_{123}]=
y_1y_2v^2[r_{12}(\lambda_1+\lambda_2) +2r(\lambda_3+\lambda_1-\lambda_2)]
\nonumber\\
d&\equiv&y_2^2v^2(2r_{12}\lambda_{123})=y_2^2v^2[2r_{12}(\lambda_3+\lambda_1-\lambda_2)].
\label{id1_d5_model1}
\end{eqnarray} 
This brings us to the completion of our first task, i.e., generating the form of the left-handed Majorana neutrino mass matrix
in Eq. (\ref{abc_d5}) corresponding to $\theta_{13}=0$, $\theta_{23}=\pi/4$ and $\theta_{12}^0$ of any of the popular lepton mixing values.
 
Next, our job is to achieve realistic neutrino mixing i.e., non-zero $\theta_{13}$, deviation of $\theta_{23}$ from $\pi/4$
and small modification of the solar mixing angle $\theta_{12}^0$ in our model. For this to happen, we will have to tinker 
the maximal mixing between the two right-handed neutrino states by a small amount or in other words 
we will have to shift from the choice of $r_{11}=r_{22}=r$ and allow the two diagonal entries of the right-handed Majorana neutrino mass matrix to differ from each other by a little amount. Thus we now set $r_{22}=r_{11}+\epsilon$, where $\epsilon$
is a small quantity. Therefore we are now again back to the most general scenario of $M_{\nu R}$ exhibiting non-maximal mixing between the two right-handed neutrino states $N_{1R}$ and $N_{2R}$ as shown in Eq. (\ref{mrh_d5}). 
Keeping $v_1=v_2=v$ condition unchanged and allowing a little shift from maximal mixing between $N_{1R}$ and $N_{2R}$ i.e.,
applying $r_{22}=r_{11}+\epsilon$, we can get the left-handed Majorana neutrino mass matrix that can be dissociated into two parts viz. a dominant part $M^0$ that appears similar to the $M_{\nu_L}^{flavour}$ in Eq. (\ref{mchoicemaxmix_d5_model1}) and a sub-dominant contribution given by $M'$ which in its turn is proportional to the tiny shift $\epsilon$: 
\begin{equation}
M_{\nu_L}^{flavour}=M^0+M'.
\label{m0_mprime}
\end{equation}   
Here\footnote{We will see in Section (\ref{secondmodel}), the $M^0$ for both Model 1 and Model 2 have the same basic structure although the $M'$ for the two models appear to be different.},
\begin{equation}
{M^0}|_{\rm{(Model\,1)}}=
v^2\pmatrix{ y_1^2[4r_{12}\lambda_{123} +2r_{11}\lambda_{12}]
& y_1y_2[r_{12}\lambda_{12} +2r_{11}\lambda_{123}]
& y_1y_2[r_{12}\lambda_{12} +2r_{11}\lambda_{123}]\cr
y_1y_2[r_{12}\lambda_{12} +2r_{11}\lambda_{123}]
& y_2^2r_{11}\lambda_{12}& y_2^2(2r_{12}\lambda_{123})\cr 
y_1y_2[r_{12}\lambda_{12} +2r_{11}\lambda_{123}]
& y_2^2(2r_{12}\lambda_{123}) & y_2^2r_{11}\lambda_{12}
},
%\, {\rm and} \, 
\label{m0_d5_model1}
\end{equation} 
and
\begin{equation}
{M'}|_{\rm{(Model\,1)}}=\epsilon \pmatrix{ x
& y & 0\cr
y
& x' & 0\cr 
0
& 0 & 0
}.
\label{mprime_d5_model1}
\end{equation} 
The $x,\, y\, {\rm and}\,x'$ in Eq. (\ref{mprime_d5_model1}) are given by:
\begin{eqnarray}
x&\equiv&y_1^2v^2\lambda_{12}
= y_1^2v^2(\lambda_1+\lambda_2)\nonumber\\
x'&\equiv& y_2^2v^2\lambda_{12}
= y_2^2v^2(\lambda_1+\lambda_2)\nonumber\\
y&\equiv&2y_1y_2v^2\lambda_{123}=
2y_1y_2v^2(\lambda_3+\lambda_1-\lambda_2).
\label{xy_d5}
\end{eqnarray} 
It is straightforward to note that $M^0$ in Eq. (\ref{m0_d5_model1}) has the same form as that of the left-handed Majorana neutrino mass matrix with $\theta_{13}=0$, $\theta_{23}=\pi/4$ and $\theta_{12}^0$ of any of the popular mixing choices as shown in Eq. (\ref{abc_d5}) if we identify:
\begin{eqnarray}
a'&\equiv&y_1^2v^2[4r_{12}\lambda_{123} +2r_{11}\lambda_{12}]
= y_1^2v^2[4r_{12}(\lambda_3+\lambda_1-\lambda_2) +2r_{11}(\lambda_1+\lambda_2)]\nonumber\\
b'&\equiv& y_2^2v^2r_{11}\lambda_{12}=y_2^2v^2r_{11}(\lambda_1+\lambda_2)\nonumber\\
c'&\equiv&y_1y_2v^2[r_{12}\lambda_{12} +2r_{11}\lambda_{123}]=
y_1y_2v^2[r_{12}(\lambda_1+\lambda_2) +2r_{11}(\lambda_3+\lambda_1-\lambda_2)]
\nonumber\\
d'&\equiv&y_2^2v^2(2r_{12}\lambda_{123})=y_2^2v^2[2r_{12}(\lambda_3+\lambda_1-\lambda_2)]
\label{id2_d5_model1}
\end{eqnarray} 
just as we did\footnote{To distinctly identify the $r_{11}=r_{22}=r$ scenario from the $r_{22}=r_{11}+\epsilon$ case, a primed notation has been used here.} for Eq. (\ref{id1_d5_model1}).

One can employ non-degenerate perturbation theory calculation techniques to obtain the corrections offered to the eigenvalues
and eigenvectors of $M^0$ by $M'$. Needless to mention that the columns of the $U^0$ mixing matrix in Eq. (\ref{mix0_d5}) represents the unperturbed flavour basis. For the equations to look less complicated we define:
\begin{equation}
 \gamma\equiv(b'-3d'-a') \ \  {\rm and} \ \ 
\rho\equiv\sqrt{a^{'2}+b^{'2}+8c^{'2}+d^{'2}-2a^{'}b^{'}-2a^{'}d^{'}+2b^{'}d^{'}}.
\label{grho_d5}
\end{equation}
The third ket after including first-order corrections looks like:
\begin{equation}
{|\psi_3\rangle}|_{\rm{(Model\,1)}}=
\pmatrix{\frac{\epsilon}{\gamma^2-\rho^2}
\left[\rho(\sqrt{2}y\cos 2\theta_{12}^0-x'\sin 2\theta_{12}^0)
-\gamma\sqrt{2}y\right]
\cr
-\frac{1}{\sqrt{2}}[1+\xi \epsilon]
\cr
\frac{1}{\sqrt{2}}[1-\xi \epsilon]}.
\label{ket3_d5_model1}
\end{equation}
The $\xi$ in Eq. (\ref{ket3_d5_model1}) is given by:
\begin{equation}
\xi\equiv[\gamma x' +\rho(x'\cos 2\theta_{12}^0+\sqrt{2}y\sin 2\theta_{12}^0)]/(\gamma^2-\rho^2).
\label{xi_d5_model1}
\end{equation}
In a CP-conserving situation:
\begin{equation}
\sin \theta_{13}=\frac{\epsilon}{\rho^2-\gamma^2}
\left[\rho(\sqrt{2}y\cos 2\theta_{12}^0-x'\sin 2\theta_{12}^0)
-\gamma\sqrt{2}y\right].
\label{s13_d5_model1}
\end{equation}
Thus one can easily get non-zero $\theta_{13}$ in terms of the model parameters such as $\epsilon$, the vev of the scalars $v$ 
and the quartic couplings $\lambda_i$, ($i=1,2,3$)
using Eqs. (\ref{id2_d5_model1}), (\ref{grho_d5}) and (\ref{s13_d5_model1}).

Eq. (\ref{ket3_d5_model1}) can also yield an expression for the deviation of $\theta_{23}$ from $\pi/4$:
\begin{equation}
\tan\varphi\equiv\tan(\theta_{23}-\pi/4)=\xi \epsilon.
\label{atmmix_d5_model1}
\end{equation}

From the first-order corrections to the first and the second ket, one can easily compute the corrections to the solar mixing
angle $\theta_{12}$:
\begin{equation}
\tan\theta_{12}=\frac{\sin\theta_{12}^0+\epsilon \beta \cos\theta_{12}^0}{\cos\theta_{12}^0-\epsilon \beta \sin\theta_{12}^0},
\label{solmix_d5_model1}
\end{equation}
with 
\begin{equation}
\beta\equiv \frac{\left[ \frac{y}{\sqrt{2}}\cos 2\theta_{12}^0
+\frac{1}{2}(x-\frac{x'}{2})\sin 2\theta_{12}^0
\right]}{\rho}.
\label{beta_d5_model1}
\end{equation}
It is obvious from Eqs. (\ref{solmix_d5_model1}) and (\ref{atmmix_d5_model1}), that one can obtain the modified $\theta_{12}$ and $\theta_{23}$ deviation from $\pi/4$ respectively in terms of the parameters of Model 1 with the help of Eqs. (\ref{id2_d5_model1}), (\ref{grho_d5}), (\ref{xi_d5_model1}) and (\ref{beta_d5_model1}).

In this whole discussion of Model 1, we have limited ourselves to the CP-conserving scenario by
keeping $r_{\alpha \beta}$, ($\alpha,\beta=1,2$) real. One can of course, in general have complex $r_{\alpha \beta}$
by assigning Majorana phases to the right-handed neutrino masses. In such a situation, $\epsilon$ will become
complex that can produce CP-violation from Eq. (\ref{ket3_d5_model1}).

\begin{figure}[tbh]
\begin{center}
\includegraphics[scale=0.22,angle=0]{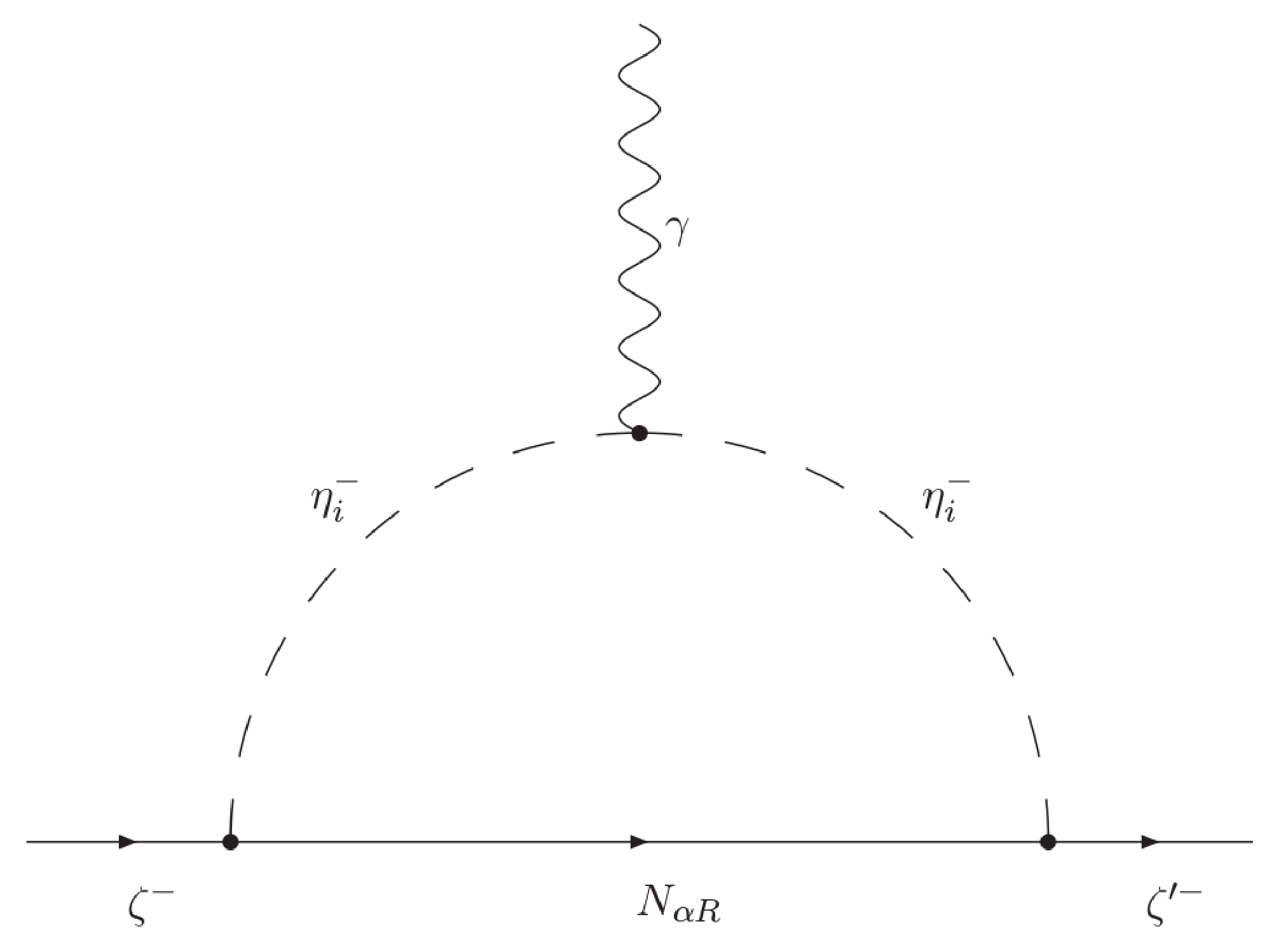}
\end{center}
\caption{\sf Diagram for charged lepton decay at one-loop for both Model 1 and Model 2. Here $\zeta^-$ and $\zeta'^-$ represent
$(e^-,\mu^-,\tau^-)$. In case of charge lepton flavour violating (LFV) decays $\zeta^- \ne \zeta'^-$.
Only the processes $\mu^-\rightarrow e^-\gamma$, 
$\tau^-\rightarrow e^-\gamma$ and $\tau^-\rightarrow \mu^-\gamma$ are kinematically allowed and thus people 
search for these decays. LFV decays at one-loop level via this diagram are forbidden by $D5$ symmetry in both 
Model 1 and Model 2.
}
\label{lfvd5} 
\end{figure}

Before closing our discussion on Model 1, let us discuss the prospects 
of flavour changing decays of charged leptons in this model.
As expected, the charged lepton flavour violation (LFV) is governed by the Yukawa Lagrangian analogous 
to that given in Eq. (\ref{yukawa_d5_model1}):
\begin{equation}
{\mathscr{L}_{{\rm LFV}}}|_{\rm{(Model\,1)}}= y_1\left[(\overline{N}_{2R}\eta_2^++\overline{N}_{1R}\eta_1^+)e^-\right]
+y_2\left[(\overline{N}_{1R}\eta_2^+)\tau^-+ (\overline{N}_{2R}\eta_1^+)\mu^-\right]+h.c.
\label{lfv_d5_model1}
\end{equation}
The LFV processes at one-loop level in Model 1 can occur through diagram\footnote{We will 
see in Section (\ref{secondmodel}) the same diagram shown in Fig. (\ref{lfvd5}) 
can also give rise to LFV decays for Model 2 at one-loop level which can be prohibited by $D5$ symmetry. Thus Fig. (\ref{lfvd5}) is valid for both Model 1 and Model 2.} as shown in Fig. (\ref{lfvd5}).
It is evident from Eq. (\ref{lfv_d5_model1}), that the kinematically allowed processes like
$\mu^-\rightarrow e^-\gamma$, 
$\tau^-\rightarrow e^-\gamma$ and $\tau^-\rightarrow \mu^-\gamma$ through Fig. (\ref{lfvd5})
is forbidden by the symmetries in the model. This is precisely due to the fact that Eq. (\ref{lfv_d5_model1}) prohibits 
the combination of the $\eta_i$ and $N_\alpha$ required at the two Yukawa vertices of Fig. (\ref{lfvd5})
to mediate LFV processes. Therefore the LFV processes at one-loop level are strictly forbidden by $D5$ symmetry
in this model.
We now conclude our analysis of Model 1 here. In the following section we will explore Model 2.

%---------------------------------------------------------------------------
%---------------------------------------------------------------------------

%--------------------
\subsection{Model 2}
\label{secondmodel}
As stated earlier in Section (\ref{Introduction_d5}), Model 2 has fields similar 
to that of Model 1 except for the fact that the $D5$ quantum
numbers of the fields in Model 2 differ from that in Model 1. 
For Model 2 also, let us denote the three left-handed lepton $SU(2)_L$ doublets contained in it 
by $L_{\zeta_L}\equiv(\nu_\zeta \ \ \zeta^-)_L^T$ with $\zeta=e,\mu,\tau$. Out of these, the two
left-handed lepton $SU(2)_L$ doublets i.e., $L_{\mu_L}$ and $L_{\tau_L}$ form a $2_2$ of $D_5$ while we assign the $D_5$ quantum number $1_1$ to $L_{e L}$. Two right-handed neutrinos, singlet of SM gauge group are also present in Model 2 that constitute a $2_1$ of $D5$. In the scalar sector of Model 2 we have two $SU(2)_L$ inert doublet scalars namely, $\eta_i\equiv(\eta_i^+, \eta_i^0)^T$, $(i=1,2)$ forming a $2_1$ of $D5$ given by $\eta$. Along with $\eta$, the scalar sector is adorned by two $SU(2)_L$ doublet scalars i.e., $\Phi_j\equiv(\phi_j^+, \phi_j^0)^T$, $(j=1,2)$ having $D5$ quantum number $2_1$ denoted by $\Phi$.
There is an unbroken $Z_2$ also present in the model along with $D5$. All the fields except the scalars $\eta$ and the two right-handed neutrinos are even under this unbroken $Z_2$. After SSB, the $Z_2$ even $\phi_j$ fields get vev but $Z_2$ odd $\eta_i$ scalars do not get any vev. Here also, the vev of $\phi_j^0$ is given by $v_j$ i.e., $\langle \Phi_j\rangle\equiv v_j$, $(j=1,2)$. All the fields in Model 2 with their corresponding quantum numbers are displayed\footnote{A glance at Table (\ref{particles_model1}) and Table (\ref{particles_model2}) will immediately reveal that the fields with $D5$ quantum number $2_2$ in Model 1 has been assigned $D5$ charge $2_1$ in Model 2 and vice versa. This in fact, is the main difference between Model 1 and Model 2.} in Table (\ref{particles_model2}). 
In Model 2 also, we delimit our study to the neutrino sector only. We work in the basis in which charged-lepton mass matrix is diagonal and the whole mixing arises from the neutrino sector. 
%--------------------------- 
\begin{table}[tb]
\begin{center}
\begin{tabular}{|c|c|c|c|}
\hline
\sf{Leptons} & $SU(2)_L$ & $D5$ & $Z_2$ \\ \hline
 & & &  \\ 
$L_{e_L}\equiv\pmatrix{\nu_e& e^-}_L$ & $2$ & $1_1$ & $1$ \\
 & & &  \\ 
 \hline
 & & &  \\ 
$L_{\zeta_L}\equiv\pmatrix{
\nu_\mu & \mu^- \cr \nu_\tau & \tau^- }_L$& $2$ & $2_2$ & $1$ \\
 & & &  \\ 
 \hline
 & & &  \\ 
$N_{\alpha R}\equiv \pmatrix{N_{1R}\cr
N_{2R}} $ & $1$ & $2_1$ & $-1$ \\ 
 & & &  \\ 
\hline
\hline
\sf{Scalars}& $SU(2)_L$ & $D5$ & $Z_2$ \\ \hline
 & & &  \\ 
$\Phi \equiv \pmatrix{\phi_1^+ & \phi_1^0\cr
\phi_2^+ & \phi_2^0 }$ & $2$ & $2_1$ & $1$ \\ 
 & & &  \\ 
\hline
 & & &  \\ 
$\eta\equiv \pmatrix{\eta_1^+ & \eta_1^0\cr
\eta_2^+ & \eta_2^0 }$ & $2$ & $2_1$ & $-1$ \\ 
 & & &  \\ 
\hline
\end{tabular}
\end{center}
\caption{\sf{Field content of Model 2 along with their respective charges. 
Our study here is delimited to the neutrino sector only.}}
\label{particles_model2}
\end{table}
%---------------------

We follow the same technique used in Model 1 to analyze Model 2.
The radiative neutrino mass at one-loop level in Model 2 can be generated by the same 
diagram as in Model 1 i.e., Fig. (\ref{radfig_d5}). From the $D5$ product rules in Appendix (\ref{D5group}),
it is evident that the total scalar potential as well as the relevant part of it viz. $V_{relevant}$ containing 
all the $(\eta^\dagger \phi)(\eta^\dagger \phi)$ terms\footnote{Since Fig. (\ref{radfig_d5}) is common to both Model 1 and Model 2, two $\eta$ are created and two $\phi$ are extirpated at the four-point scalar vertex causing
$(\eta^\dagger \phi)(\eta^\dagger \phi)$ terms of the scalar potential to be the one relevant for the analysis in Model 2 also.} that regulate the structure of the left-handed Majorana neutrino mass matrix in Model 2 are the same as in Model 1. Therefore the $V_{relevant}$ for Model 2 is given by Eq. (\ref{potential_d5}) mentioned in the previous section\footnote{In other words, in spite of the differences between the $D5$ quantum numbers of the scalars in Model 1 and Model 2, from the product rules in Appendix (\ref{D5group}) it can be easily perceived that the scalar potential for Model 1 and Model 2 are exactly identical as discussed in Appendix (\ref{scalarpotentials}).}.

As usual, all symmetries are conserved at all the three vertices we have in Fig. (\ref{radfig_d5}).
The $D5\times Z_2$ conserving Dirac vertices can be written as:
\begin{equation}
{\mathscr{L}_{Yukawa}}|_{\rm{(Model\,2)}}= \widetilde{y_1}\left[(\overline{N}_{2R}\eta_2^0+\overline{N}_{1R}\eta_1^0)\nu_e\right]
+\widetilde{y_2}\left[(\overline{N}_{2R}\eta_1^0)\nu_\tau+ (\overline{N}_{1R}\eta_2^0)\nu_\mu\right]+h.c.
\label{yukawa_d5_model2}
\end{equation}
As the $D_5$ quantum numbers of the left-handed neutrinos $\nu_\zeta$ for ($\zeta=\mu,\,\tau$) is $2_2$ whereas that of 
$\nu_e$ is $1_1$, we have different Yukawa couplings for them denoted by $\widetilde{y_2}$ and $\widetilde{y_1}$ respectively in Eq. (\ref{yukawa_d5_model2}). It is also worth pointing out that owing to the difference between the $D5$ charges of the fields
in Model 1 and Model 2, we assign different names to the Yukawa couplings of Model 2 
in Eq. (\ref{yukawa_d5_model2}) i.e., $\widetilde{y_i}$, ($i=1,2$) compared to the Yukawa couplings of Model 1 
given by $y_i$, ($i=1,2$) in Eq. (\ref{yukawa_d5_model1}). 
It must also be noted that the $\widetilde{y_2}$ term in Eq. (\ref{yukawa_d5_model2}) of Model 2 
looks substantially different from the $y_2$ term Eq. (\ref{yukawa_d5_model1}) of Model 1 because of 
the difference between $D5$ charges of the fields in Model 1 and Model 2.

In the right-handed neutrino sector of Model 2, essentially one can follow the same steps as in Model 1
to achieve the right-handed Majorana neutrino mass matrix as in Eq. (\ref{mrh_d5}).
Although the $D5$ charge of the right-handed neutrino fields is $2_1$ in Model 2 whereas it is $2_2$ in Model 1,
from the product rules in Appendix (\ref{D5group}), it is easy to comprehend that one can 
achieve exactly the same direct mass term as in Eq. (\ref{rhnu_model1_d5intact}), allow the same soft $D5$ breaking terms as in Eq. (\ref{rhnu_model1_d5softbreak}) and finally obtain the same right-handed neutrino mass matrix $M_{\nu R}$ as in Eq. (\ref{mrh_d5}) for Model 2 as well.

As there is no difference in the essential inputs required to tackle the dark matter part as well as the $Z_2$ charges of the fields in Model 1 and Model 2, we can handle the dark matter sector of Model 2 exactly in the same way\footnote{Since 
one can replicate the same assumptions and steps in Model 2 in analogy to Model 1, we are not repeating the discussion about the dark matter candidate here again. One can refer to the relevant part of Section (\ref{firstmodel}) to explore it in more details.} as 
we did in Model 1 and identify the lightest of the $\eta_i$, ($i=1,2$) fields to be the dark matter component of Model 2 as we did for Model 1.

The next step in the procedure is to discuss the contributions coming from 
the one-loop diagram in Fig. (\ref{radfig_d5}) to
the left-handed Majorana neutrino mass matrix in Model 2. 
For this purpose, we follow similar logical steps and make 
similar assumptions one by one as we did in Section (\ref{firstmodel}) for Model 1 
with necessary alterations required for Model 2 taken into account\footnote{We are 
not discussing each step and are just writing down 
the main equations for Model 2 here. One can refer to the 
particular discussion available in Section (\ref{firstmodel}) for Model 1 as and when needed.}. 
In the similar fashion we can define $z\equiv \frac{m_R^2}{m_0^2}$
and write down the second diagonal entry of $M_{\nu_L}^{flavour}$ in the limit $m_R^2>>m_0^2$ as\footnote{For
complete insight about the factors $m_R$, $m_0$, $z$, $\lambda$ and the associated assumptions required to deduce them, 
refer to respective discussion in Section (\ref{firstmodel}) 
in context of Model 1. The same remains valid for Model 2 also. }:
\begin{equation}
{(M_{\nu_L}^{flavour})_{22}}|_{\rm{(Model\,2)}}=\lambda \frac{v_m v_n}{8 \pi^2}
\frac{(\widetilde{y_2})^2 }{m_{R_{11}}} 
\left[ \ln z -1 \right].
\label{m22_d5_model2}
\end{equation}
Note that the RHS of Eq. (\ref{m22_d5_model2}) contains $m_{R_{11}}$ whereas the RHS of 
Eq. (\ref{m22_d5_model1}) has $m_{R_{22}}$. This is primarily due to the difference between the $\widetilde{y_2}$ term
of Eq. (\ref{yukawa_d5_model2}) and $y_2$ term of Eq. (\ref{yukawa_d5_model1}) which is the manifestation of 
the differences between the $D5$ quantum numbers of the fields in Model 2 and Model 1. 
Keeping in mind the essential features of the $\widetilde{y_2}$ term of Eq. (\ref{yukawa_d5_model2}), 
one can obtain the the expression for $(M_{\nu_L}^{flavour})_{33}$ for Model 2 by simply replacing 
$m_{R_{11}}$ of Eq. (\ref{m22_d5_model2}) by $m_{R_{22}}$, in an analogous manner as we did for Model 1 in Section (\ref{firstmodel}).

Applying similar logic as for Eq. (\ref{m23_d5_Model1}) of Model 1
in Section (\ref{firstmodel}) and using the $\widetilde{y_2}$ term of Eq. (\ref{yukawa_d5_model2}) together 
with the mass insertion approximation, we can calculate the off-diagonal entry $(M_{\nu_L}^{flavour})_{23}$ for Model 2 also:
\begin{equation}
{(M_{\nu_L}^{flavour})_{23}}|_{\rm{(Model\,2)}}=\lambda \frac{v_m v_n}{8 \pi^2}
\frac{(\widetilde{y_2})^2 m_{R_{12}}}{m_{R_{11}}m_{R_{22}}} 
\left[ \ln z -1 \right].
\label{m23_d5_Model2}
\end{equation} 
Expressions for the rest of the entries of $M_{\nu_L}^{flavour}$ such as the (1,1), (1,2) and (1,3) elements can be deduced 
in a similar way. 

We can absorb all the entities in the RHS of Eqs. (\ref{m22_d5_model2}) 
and (\ref{m23_d5_Model2}) except $\lambda$, $\widetilde{y_i}$ and $v_i$ 
in some loop-contributing factors just as 
we did in Eq. (\ref{rij_d5}) of Section (\ref{firstmodel}) for Model 1. 
In fact, we can use the same definitions of $r_{\alpha\beta}$ as in Eq. (\ref{rij_d5}) for Model 2 also.

Utilizing Eqs. (\ref{m22_d5_model2}), (\ref{m23_d5_Model2}), (\ref{rij_d5}) and (\ref{potential_d5}) we can
write $M_{\nu_L}^{flavour}$ for Model 2 just as we did in Eq. (\ref{mgeneral_d5_model1}) for Model 1 as:
\begin{equation}
{M_{\nu_L}^{flavour}}|_{\rm{(Model\,2)}}=\pmatrix{ \widetilde{\chi}_1
& \widetilde{\chi}_4 & \widetilde{\chi}_5 \cr
\widetilde{\chi}_4
& \widetilde{\chi}_2 & \widetilde{\chi}_6 \cr 
\widetilde{\chi}_5
& \widetilde{\chi}_6 & \widetilde{\chi}_3
}.
\label{mgeneral_d5_model2}
\end{equation}
The $\widetilde{\chi}_k$ with ($k=1,2,3,...,6$) in Eq. (\ref{mgeneral_d5_model2}) is given by:
\begin{eqnarray}
\widetilde{\chi}_1&\equiv&(\widetilde{y_1})^2\left[
4 r_{12} v_1v_2(\lambda_3+\lambda_1-\lambda_2)
+(r_{11}v_1^2+r_{22}v_2^2)(\lambda_1+\lambda_2)
\right]\nonumber\\
\widetilde{\chi}_2&\equiv&(\widetilde{y_2})^2\left[r_{11}(\lambda_1+\lambda_2)v_2^2\right]\nonumber\\
\widetilde{\chi}_3&\equiv&(\widetilde{y_2})^2\left[r_{22}(\lambda_1+\lambda_2)v_1^2\right]\nonumber\\
\widetilde{\chi}_4&\equiv& \widetilde{y_1}\widetilde{y_2}\left[
r_{12}(\lambda_1+\lambda_2)v_2^2+2r_{11}(\lambda_3+\lambda_1-\lambda_2)v_1v_2
\right]\nonumber\\
\widetilde{\chi}_5&\equiv& 
\widetilde{y_1}\widetilde{y_2}\left[
r_{12}(\lambda_1+\lambda_2)v_1^2+2r_{22}(\lambda_3+\lambda_1-\lambda_2)v_1v_2
\right]
\nonumber\\
\widetilde{\chi}_6&\equiv& (\widetilde{y_2})^2\left[
2 r_{12}(\lambda_3+\lambda_1-\lambda_2)v_1v_2
\right], 
\label{chidefinition_d5_model2}
\end{eqnarray}
where, $\langle \Phi_j\rangle\equiv v_j$ with ($j=1,2$).

To get the form of the left-handed Majorana neutrino mass matrix as in Eq. (\ref{abc_d5}) corresponding to 
$\theta_{13}=0$, $\theta_{23}=\pi/4$ and $\theta_{12}^0$ of any of the popular mixing values given in Table (\ref{tab1}),
we need to have $\widetilde{\chi}_1\ne\widetilde{\chi}_2=\widetilde{\chi}_3$ 
along with $\widetilde{\chi}_4=\widetilde{\chi}_5$ in Eq. (\ref{mgeneral_d5_model2}). This can be achieved by
setting $v_1=v_2=v$ and $r_{11}=r_{22}=r$ in Eq. (\ref{chidefinition_d5_model2}) just in the same way as we did
for Model 1 in Section (\ref{firstmodel}). The right-handed neutrino mass matrix once again boils down to the 
exactly same one as shown in Eq. (\ref{mrh2_d5}) that corresponds to the maximal mixing between the 
two right-handed neutrinos $N_{1R}$ and $N_{2R}$ as we have applied $r_{11}=r_{22}=r$ condition. Thus the right-handed neutrino
mass matrix after application of $r_{11}=r_{22}=r$ condition in Model 2 is same as that in Model 1 and is given by Eq. (\ref{mrh2_d5}). Implementing the conditions $v_1=v_2=v$ and $r_{11}=r_{22}=r$ in Eq. (\ref{mgeneral_d5_model2}) we get:
\begin{equation}
{M_{\nu_L}^{flavour}}|_{\rm{(Model\,2)}}=v^2\pmatrix{ (\widetilde{y_1})^2[4r_{12}\lambda_{123} +2r\lambda_{12}]
& \widetilde{y_1}\widetilde{y_2}[r_{12}\lambda_{12} +2r\lambda_{123}]
& \widetilde{y_1}\widetilde{y_2}[r_{12}\lambda_{12} +2r\lambda_{123}]\cr
\widetilde{y_1}\widetilde{y_2}[r_{12}\lambda_{12} +2r\lambda_{123}]
&(\widetilde{y_2})^2r\lambda_{12}& (\widetilde{y_2})^2(2r_{12}\lambda_{123})\cr 
\widetilde{y_1}\widetilde{y_2}[r_{12}\lambda_{12} +2r\lambda_{123}]
& (\widetilde{y_2})^2(2r_{12}\lambda_{123}) & (\widetilde{y_2})^2r\lambda_{12}
}.
\label{mchoicemaxmix_d5_model2}
\end{equation} 
Recall, $\lambda_{12}\equiv\lambda_1+\lambda_2$ and $\lambda_{123}\equiv\lambda_3+\lambda_1-\lambda_2$.
The following mappings are needed to visualize the $M_{\nu_L}^{flavour}$
in Eq. (\ref{mchoicemaxmix_d5_model2}) to be of the form as mentioned in Eq. (\ref{abc_d5}):
\begin{eqnarray}
a&\equiv&(\widetilde{y_1})^2v^2[4r_{12}\lambda_{123} +2r\lambda_{12}]
= (\widetilde{y_1})^2v^2[4r_{12}(\lambda_3+\lambda_1-\lambda_2) +2r(\lambda_1+\lambda_2)]\nonumber\\
b&\equiv& (\widetilde{y_2})^2v^2r\lambda_{12}=(\widetilde{y_2})^2v^2r(\lambda_1+\lambda_2)\nonumber\\
c&\equiv&\widetilde{y_1}\widetilde{y_2}v^2[r_{12}\lambda_{12} +2r\lambda_{123}]=
\widetilde{y_1}\widetilde{y_2}v^2[r_{12}(\lambda_1+\lambda_2) +2r(\lambda_3+\lambda_1-\lambda_2)]
\nonumber\\
d&\equiv&(\widetilde{y_2})^2v^2(2r_{12}\lambda_{123})=y_2^2v^2[2r_{12}(\lambda_3+\lambda_1-\lambda_2)].
\label{id1_d5_model2}
\end{eqnarray}

To get the realistic neutrino mixings, i.e., 
non-zero $\theta_{13}$, deviation of $\theta_{23}$ from $\pi/4$
and tiny modification of the solar mixing angle $\theta_{12}^0$ in Model 2
one requires to shift from maximal 
mixing between the two right-handed neutrino states by a small amount 
$\epsilon$ i.e., by applying $r_{22}=r_{11}+\epsilon$ with $v_1=v_2=v$.
As we discussed for Model 1, here also, applying the condition $r_{22}=r_{11}+\epsilon$ 
basically causes the right-handed neutrino mass matrix to resume its general form with 
unequal diagonal entries as in Eq. (\ref{mrh_d5}). 
Putting $r_{22}=r_{11}+\epsilon$ in Eq. (\ref{mchoicemaxmix_d5_model2}) with 
$v_1=v_2=v$ condition still valid we can write $M_{\nu_L}^{flavour}=M^0+M'$
as we did for Model 1 in Eq. (\ref{m0_mprime}). The $M^0$ is the dominant contribution and the $M'$ is the subdominant contribution proportional to $\epsilon$ that can be expressed as below:
\begin{equation}
{M^0}|_{\rm{(Model\,2)}}=
v^2\pmatrix{ (\widetilde{y_1})^2[4r_{12}\lambda_{123} +2r_{11}\lambda_{12}]
& \widetilde{y_1}\widetilde{y_2}[r_{12}\lambda_{12} +2r_{11}\lambda_{123}]
& \widetilde{y_1}\widetilde{y_2}[r_{12}\lambda_{12} +2r_{11}\lambda_{123}]\cr
\widetilde{y_1}\widetilde{y_2}[r_{12}\lambda_{12} +2r_{11}\lambda_{123}]
& (\widetilde{y_2})^2r_{11}\lambda_{12}& (\widetilde{y_2})^2(2r_{12}\lambda_{123})\cr 
\widetilde{y_1}\widetilde{y_2}[r_{12}\lambda_{12} +2r_{11}\lambda_{123}]
& (\widetilde{y_2})^2(2r_{12}\lambda_{123}) & (\widetilde{y_2})^2r_{11}\lambda_{12}
},
\label{m0_d5_model2}
\end{equation} 
and
\begin{equation}
{M'}|_{\rm{(Model\,2)}}=\epsilon \pmatrix{ \widetilde{x}
& 0 & \widetilde{y}\cr
0
& 0 & 0\cr 
\widetilde{y}
& 0 & \widetilde{x}'
}.
\label{mprime_d5_model2}
\end{equation} 
The $\widetilde{x},\, \widetilde{y}\, {\rm and}\,\widetilde{x}'$ 
mentioned in Eq. (\ref{mprime_d5_model1}) can be written as:
\begin{eqnarray}
\widetilde{x}&\equiv&(\widetilde{y_1})^2v^2\lambda_{12}
= (\widetilde{y_1})^2v^2(\lambda_1+\lambda_2)\nonumber\\
\widetilde{x}'&\equiv& (\widetilde{y_2})^2v^2\lambda_{12}
= (\widetilde{y_2})^2v^2(\lambda_1+\lambda_2)\nonumber\\
\widetilde{y}&\equiv&2\widetilde{y_1}\widetilde{y_2}v^2\lambda_{123}=
2\widetilde{y_1}\widetilde{y_2}v^2(\lambda_3+\lambda_1-\lambda_2).
\label{xy_d5_model2}
\end{eqnarray} 
In order to ascertain that the $M^0$ in Eq. (\ref{m0_d5_model2}) basically has the same
form as that of the left-handed Majorana neutrino mass matrix specific to $\theta_{13}=0$, 
$\theta_{23}=\pi/4$ and $\theta_{12}^0$ of the popular mixing kind as shown in Eq. (\ref{abc_d5}), 
one has to make the following identifications\footnote{We applied
the same procedure for Model 1 while obtaining Eq. (\ref{id2_d5_model1}).}: 
\begin{eqnarray}
\widetilde{a'}&\equiv&(\widetilde{y_1})^2v^2[4r_{12}\lambda_{123} +2r_{11}\lambda_{12}]
= (\widetilde{y_1})^2v^2[4r_{12}(\lambda_3+\lambda_1-\lambda_2) +2r_{11}(\lambda_1+\lambda_2)]\nonumber\\
\widetilde{b}'&\equiv& (\widetilde{y_2})^2v^2r_{11}\lambda_{12}=(\widetilde{y_2})^2v^2r_{11}(\lambda_1+\lambda_2)\nonumber\\
\widetilde{c}'&\equiv&\widetilde{y_1}\widetilde{y_2}v^2[r_{12}\lambda_{12} +2r_{11}\lambda_{123}]=
\widetilde{y_1}\widetilde{y_2}v^2[r_{12}(\lambda_1+\lambda_2) +2r_{11}(\lambda_3+\lambda_1-\lambda_2)]
\nonumber\\
\widetilde{d}'&\equiv&(\widetilde{y_2})^2v^2(2r_{12}\lambda_{123})=(\widetilde{y_2})^2v^2[2r_{12}(\lambda_3+\lambda_1-\lambda_2)]
\label{id2_d5_model2}
\end{eqnarray} 
Now we apply non-degenerate perturbation theory to calculate the corrections to $M^0$ coming from $M'$.
Once again, we remind ourselves that the columns of the $U^0$ mixing matrix in Eq. (\ref{mix0_d5}) are the unperturbed
flavour basis.
The third ket after getting first order corrections is given by:
\begin{equation}
{|\psi_3\rangle}|_{\rm{(Model\,2)}} =
\pmatrix{-\frac{\epsilon}{\widetilde{\gamma}^2-\widetilde{\rho}^2}
\left[\widetilde{\rho}(\sqrt{2}\widetilde{y}\cos 2\theta_{12}^0-\widetilde{x}'\sin 2\theta_{12}^0)
-\widetilde{\gamma}\sqrt{2}\widetilde{y}\right]
\cr
-\frac{1}{\sqrt{2}}[1-\widetilde{\xi} \epsilon]
\cr
\frac{1}{\sqrt{2}}[1+\widetilde{\xi} \epsilon]}.
\label{ket3_d5_model2}
\end{equation}
where, 
\begin{equation}
\widetilde{\gamma}\equiv(\widetilde{b}'-3\widetilde{d}'-\widetilde{a'}) \ \  {\rm and} \ \ 
\widetilde{\rho}\equiv\sqrt{\widetilde{a^{'2}}+\widetilde{b}^{'2}+8\widetilde{c}^{'2}+\widetilde{d}^{'2}
-2\widetilde{a^{'}}\widetilde{b}^{'}-2\widetilde{a^{'}}\widetilde{d}^{'}+2\widetilde{b}^{'}\widetilde{d}^{'}},
\label{grho_d5_model2}
\end{equation}
and 
\begin{equation}
\widetilde{\xi}\equiv[\widetilde{\gamma} \widetilde{x}' +\widetilde{\rho}(\widetilde{x}'\cos 2\theta_{12}^0
+\sqrt{2}\widetilde{y}\sin 2\theta_{12}^0)]/(\widetilde{\gamma}^2-\widetilde{\rho}^2).
\label{xi_d5_model2}
\end{equation}
%------------------------
For a CP-conserving situation:
\begin{equation}
\sin \theta_{13}=-\frac{\epsilon}{\widetilde{\rho}^2-\widetilde{\gamma}^2}
\left[\widetilde{\rho}(\sqrt{2}\widetilde{y}\cos 2\theta_{12}^0-\widetilde{x}'\sin 2\theta_{12}^0)
-\widetilde{\gamma}\sqrt{2}\widetilde{y}\right].
\label{s13_d5_model2}
\end{equation}
It is straightforward to obtain the non-zero $\theta_{13}$ in terms of Model 2 parameters such as $\epsilon$,
the vev of the scalars $v$ and $\lambda_i$ ($i=1,2,3$) using Eqs. (\ref{id2_d5_model2}), (\ref{grho_d5_model2}) and
(\ref{s13_d5_model2}). Note the difference in the non-zero $\theta_{13}$ yielded by Model 1 in Eq. (\ref{s13_d5_model1})
and that from Model 2 in Eq. (\ref{s13_d5_model2}).

The deviation of $\theta_{23}$ from $\pi/4$ can also be calculated from Eq. (\ref{s13_d5_model2}):
\begin{equation}
-\tan\widetilde{\varphi}\equiv\tan(\theta_{23}-\pi/4)=-\widetilde{\xi} \epsilon.
\label{atmmix_d5_model2}
\end{equation}
The shift of $\theta_{23}$ from $\pi/4$ given by Model 2 in Eq. (\ref{atmmix_d5_model2}) is different from
that given by Model 1 in Eq. (\ref{atmmix_d5_model1}).

The modifications of $\theta_{12}$ can be obtained from the second ket and first ket after including 
first-order corrections into them as:
\begin{equation}
\tan\theta_{12}=\frac{\sin\theta_{12}^0+\epsilon \widetilde{\beta} \cos\theta_{12}^0}{\cos\theta_{12}^0-\epsilon \widetilde{\beta} \sin\theta_{12}^0}.
\label{solmix_d5_model2}
\end{equation}
The $\widetilde{\beta}$ in Eq. (\ref{solmix_d5_model2}) is given by:
\begin{equation}
\widetilde{\beta}\equiv \frac{\left[ \frac{\widetilde{y}}{\sqrt{2}}\cos 2\theta_{12}^0
+\frac{1}{2}(\widetilde{x}-\frac{\widetilde{x}'}{2})\sin 2\theta_{12}^0
\right]}{\widetilde{\rho}}.
\label{beta_d5_model2}
\end{equation}
Although the solar mixing in Eq. (\ref{solmix_d5_model1}) coming from 
Model 1 looks very similar to that given by Model 2 in Eq. (\ref{solmix_d5_model2}),
the main difference lies in the fact that altogether $\beta$ of Model 1 in Eq. (\ref{beta_d5_model1})
is different from $\widetilde{\beta}$ of Model 2 in Eq. (\ref{beta_d5_model2}).

It is trivial to express the corrected solar mixing in Eq. (\ref{solmix_d5_model2}) and deviation of atmospheric mixing $\theta_{23}$ from $\pi/4$ in Eq. (\ref{atmmix_d5_model2}) in terms of parameters 
of Model 2 with help of Eqs. (\ref{id2_d5_model2}), (\ref{grho_d5_model2}),
(\ref{xi_d5_model2}) and (\ref{beta_d5_model2}).

A CP-conserving scenario has been studied through-out by keeping $r_{\alpha\beta}$, ($\alpha, \beta =1,2$) real.
If one considers complex $r_{\alpha\beta}$ by assigning Majorana phases to right-handed neutrinos, then $\epsilon$
will develop a complex nature which can give rise to CP-violation from Eq. (\ref{ket3_d5_model2}).

In an exactly similar fashion as in Model 1, we can show that 
LFV at one-loop level is forbidden by $D5$ symmetry in Model 2
also. The Yukawa Lagrangian responsible for charged lepton flavour violation in Model 2:
\begin{equation}
{\mathscr{L}_{{\rm LFV}}}|_{\rm{(Model\,2)}}= \widetilde{y_1}\left[(\overline{N}_{2R}\eta_2^++\overline{N}_{1R}\eta_1^+)e^-\right]
+\widetilde{y_2}\left[(\overline{N}_{2R}\eta_1^+)\tau^-+ (\overline{N}_{1R}\eta_2^+)\mu^-\right]+h.c.
\label{lfv_d5_model2}
\end{equation}
The one-loop diagram for LFV decays viz. Fig. (\ref{lfvd5}) shown in Section (\ref{firstmodel}) is valid for Model 2 also.
As already discussed in Section (\ref{firstmodel}), kinematically allowed LFV decays such as
$\mu^-\rightarrow e^-\gamma$, 
$\tau^-\rightarrow e^-\gamma$ and $\tau^-\rightarrow \mu^-\gamma$ at one-loop level in Model 2 cannot 
take place through Fig. (\ref{lfvd5}) as the combinations of $\eta_i$ and $N_{\alpha}$ needed at the two Yukawa
vertices of Fig. (\ref{lfvd5}) for these LFV processes are forbidden by Eq. (\ref{lfv_d5_model2}).
Thus both in Model 1 and in Model 2, LFV decays at one-loop are prohibited by the $D5$ symmetry.

\section{Conclusions}
\label{conclusions}
In this paper, we have devised a mechanism of scotogenic generation of realistic neutrino mixing at one-loop level 
with $D5\times Z_2$ symmetry. We demonstrate this mechanism in two set-ups governed by $D5\times Z_2$ symmetry
viz. Model 1 and Model 2. In the two models, similar fields are present with different $D5$ charges. 
In both models, two right-handed neutrinos $N_{1R}$ and $N_{2R}$ are present that can be maximally mixed to 
get the form of left-handed Majorana neutrino mass matrix corresponding to $\theta_{13}=0$, 
$\theta_{23}=\pi/4$ and any value of $\theta_{12}^0$ particular to the Tribimaximal (TBM), Bimaximal (BM), Golden Ratio (GR) or other mixings. Minute tinkering with the maximal mixing between $N_{1R}$ and $N_{2R}$ can yield realistic mixings such as
non-zero $\theta_{13}$, deviation of $\theta_{23}$ from $\pi/4$
and small modification of the solar mixing angle $\theta_{12}^0$ in a single stroke 
for both the models. In both Model 1 and Model 2,
two $Z_2$ odd inert $SU(2)_L$ doublet scalars $\eta_i$ ($i=1,2$) are present. The lightest of these two $\eta_i$ 
is a suitable dark matter candidate for both Model 1 and Model 2.

%------------------------

%===================================
{\bf Acknowledgements:} I thank Prof. Amitava Raychaudhuri for useful discussions.

%=======================================
%%%%%%%%%%%%%%%%%%%%%%%%%%%%%%%%%%%% Acknowledgements %%%%%%%%%%%%%%%%%%%%%%%%%%%%%%%%%%%%%%%%%%
\renewcommand{\thesection}{\Alph{section}} 
\setcounter{section}{0} 
\renewcommand{\theequation}{\thesection.\arabic{equation}}

\setcounter{equation}{0}

\section{Appendix: The $D5$ group}
\label{D5group}
Here we present a brief summary of the discrete group $D5$. A detailed description can be found in
\cite{Ma_d5, Lindner_d5} and \cite{Ishimori_d5}.
The discrete dihedral symmetry $D5$ basically corresponds to the symmetry of a regular pentagon.
The generators of $D5$ are $\mathcal{X}$ and $\mathcal{Y}$. The group is generated by $2\pi/5$ rotation 
$\mathcal{X}$ and reflection $\mathcal{Y}$. The generators satisfy the following relations: $\mathcal{X}^5=I$, $\mathcal{Y}^2=I$
and $\mathcal{Y}\mathcal{X}\mathcal{Y}=I$. The group $D5$ has 10 elements and there are four conjugacy classes. 
$D5$ has four irreducible representations viz. two one-dimensional representations denoted by $1_1$ and $1_2$ and 
two two-dimensional representations $2_1$ and $2_2$.

The product rules for $D5$ are as follows:
\begin{eqnarray}
1_1\,(a) \ \ \times \ \  1_1\,(b)\ \ = \ \ 1_2\,(a) \ \ \times \ \  1_2\,(b) &=& 1_1(ab),\nonumber\\
1_1\,(a) \ \ \times \ \  1_2\,(b)\ \ = \ \ 1_2\,(a) \ \ \times \ \  1_1\,(b) &=& 1_2(ab),
\label{d5product_set1}
\end{eqnarray}
\begin{eqnarray}
1_1\,(a)\ \ \times \ \  2_k\,\pmatrix{b_1\cr b_2}&=& 2_k\,\pmatrix{ab_1\cr ab_2},\nonumber\\
1_2\,(a)\ \ \times \ \  2_k\,\pmatrix{b_1\cr b_2}&=& 2_k\,\pmatrix{ab_1\cr -ab_2},\ \ (k\,=\,1,\, 2),
\label{d5product_set2}
\end{eqnarray}
\begin{eqnarray}
2_1\,\pmatrix{a_1\cr a_2}\ \ \times\ \ 2_1\,\pmatrix{b_1\cr b_2}
&=& 1_1\,(a_1b_2+a_2b_1)\ \ + \ \ 1_2\,(a_1b_2-a_2b_1)\ \ + \ \ 2_2\,\pmatrix{a_1b_1\cr a_2b_2},\nonumber\\
2_2\,\pmatrix{a_1\cr a_2}\ \ \times\ \ 2_2\,\pmatrix{b_1\cr b_2}
&=& 1_1\,(a_1b_2+a_2b_1)\ \ + \ \ 1_2\,(a_1b_2-a_2b_1)\ \ + \ \ 2_1\,\pmatrix{a_2b_2\cr a_1b_1},\nonumber\\
2_1\,\pmatrix{a_1\cr a_2}\ \ \times\ \ 2_2\,\pmatrix{b_1\cr b_2}
&=& 2_1\,\pmatrix{a_2b_1\cr a_1b_2} \ \ + \ \ 2_2\,\pmatrix{a_2b_2\cr a_1b_1}.
\label{d5product_set2}
\end{eqnarray}
These product rules play a crucial role in the models discussed in this paper.

%----------------------------
\setcounter{equation}{0}
\section{Appendix: The scalar potentials of both the models:}
\label{scalarpotentials}
The scalar sectors of Model 1 and Model 2 are present in 
Table (\ref{particles_model1}) and Table (\ref{particles_model2}) respectively.
For Model 1, we have two $SU(2)_L$ doublet scalars $\Phi_j\equiv(\phi_j^+, \phi_j^0)^T$, $(j=1,2)$
that transform as $2_2$ of $D5$ denoted by $\Phi$ having $Z_2$ charge $+1$.  In addition to that, Model 1 also has a couple of inert $SU(2)_L$ doublet scalars $\eta$, given by $\eta_i\equiv(\eta_i^+, \eta_i^0)^T$, $(i=1,2)$ that transform as $2_2$ of $D5$
and the $Z_2$ quantum number of $\eta$ is $-1$.

The picture is slightly different for Model 2. In the scalar sector of Model 2, the two $SU(2)_L$ doublet scalars 
$\Phi_j\equiv(\phi_j^+, \phi_j^0)^T$, $(j=1,2)$, represented by $\Phi$, transform as $2_1$ of $D5$ with $Z_2$ charge $+1$
whereas the inert $SU(2)_L$ doublet scalars $\eta_i\equiv(\eta_i^+, \eta_i^0)^T$, $(i=1,2)$, compactly denoted by $\eta$, transform as $2_1$ of $D5$. In Model 2 also, $\eta$ is odd under $Z_2$.

For both Model 1 and Model 2 since $\Phi$ is even under $Z_2$ it can get vev after SSB \footnote{As mentioned earlier, the vevs of $\phi_j^0$ is given by $v_j$ i.e., $\langle \Phi_j\rangle\equiv v_j$, $(j=1,2)$.} whereas $\eta$ being $Z_2$ odd does not get vev.

From the $D5$ product rules mentioned in Appendix (\ref{D5group}), one can easily infer that despite 
of the difference in the $D5$ charges of $\Phi$ and $\eta$ in Model 1 and Model 2, 
the complete scalar potential containing all terms permitted by the SM gauge 
symmetry as well as $D5\times Z_2$ are same in both the models. 
The total scalar potential inclusive of all terms allowed by the SM gauge symmetry and $D5\times Z_2$ for both
Model 1 and Model 2:
\begin{eqnarray}
V_{total}&=& m^2_\eta \left(\eta_2^\dagger\eta_2+\eta_1^\dagger\eta_1\right)
+m^2_\phi \left(\phi_2^\dagger\phi_2+\phi_1^\dagger\phi_1\right)\nonumber\\
&+& \widehat{\lambda}_1\left(\eta_2^\dagger\eta_2+\eta_1^\dagger\eta_1\right)^2
+\widehat{\lambda}_2\left(\eta_2^\dagger\eta_2-\eta_1^\dagger\eta_1\right)^2
+ \widehat{\lambda}_3 \left(\phi_2^\dagger\phi_2+\phi_1^\dagger\phi_1\right)^2
+\widehat{\lambda}_4 \left(\phi_2^\dagger\phi_2-\phi_1^\dagger\phi_1\right)^2\nonumber\\
&+&\widehat{\lambda}_5\left[\left(\eta_2^\dagger\eta_2+\eta_1^\dagger\eta_1\right)
\left(\phi_2^\dagger\phi_2+\phi_1^\dagger\phi_1\right)\right]
+\widehat{\lambda}_6\left[\left(\eta_2^\dagger\eta_2-\eta_1^\dagger\eta_1\right)
\left(\phi_2^\dagger\phi_2-\phi_1^\dagger\phi_1\right)\right]\nonumber\\
&+&\widehat{\lambda}_{7}\left[\left( \phi_1^\dagger\phi_2\right)\left( \phi_2^\dagger\phi_1\right)\right]
+\widehat{\lambda}_{8}\left[\left( \eta_1^\dagger\eta_2\right)\left( \eta_2^\dagger\eta_1\right)\right]\nonumber\\
&+&\widehat{\lambda}_{9}\left[\left\{\left( \phi_1^\dagger\phi_2\right)\left( \eta_2^\dagger\eta_1\right)\right  \} 
+ \left\{\left( \phi_2^\dagger\phi_1\right)\left( \eta_1^\dagger\eta_2\right)\right  \} 
\right] + V_{relevant} 
\label{totalpotential_d5}
\end{eqnarray}
with,
\begin{eqnarray}
V_{relevant}&=& \lambda_1 \left[ \left\{(\eta_2^\dagger \phi_2+\eta_1^\dagger \phi_1)^2 \right  \}  + h.c.\right]+ \lambda_2\left[ \left\{(\eta_2^\dagger \phi_2- \eta_1^\dagger \phi_1 )^2 \right \} +h.c.
\right]\nonumber\\
&+&\lambda_3\left[ \left\{(\eta_1^\dagger \phi_2)(\eta_2^\dagger \phi_1)+(\eta_2^\dagger \phi_1)(\eta_1^\dagger \phi_2)\right  \} 
+h.c.\right].
\label{potential2_d5}
\end{eqnarray}
%==========================================
As already discussed earlier, two $\eta$ scalars are created and two $\phi$ fields are obliterated at the four-point scalar vertex of Fig. (\ref{radfig_d5}). Thus only the $(\eta^\dagger \phi)(\eta^\dagger \phi)$ terms of the scalar potential 
take part in determining the left-handed Majorana neutrino mass matrix and hence are relevant for our purpose.
We accumulate all such $(\eta^\dagger \phi)(\eta^\dagger \phi)$ terms together in Eq. (\ref{potential2_d5}) and call it $V_{relevant}$. The couplings $\lambda_j$ ($j=1,2,3$) in Eq. (\ref{potential2_d5}) were taken to be real throughout.

%===============================


\begin{thebibliography}{100} 
% 100 is a random guess of the total number of
%references
%******************************************************************
\bibitem{t13_d5}
For the present status of $\theta_{13}$ see presentations from
Double Chooz, RENO, Daya Bay, and T2K at Neutrino
2016 (http://neutrino2016.iopconfs.org/programme).
%----



%\cite{Esteban:2020cvm}
\bibitem{nufit2022_d5}
I.~Esteban, M.~C.~Gonzalez-Garcia, M.~Maltoni, T.~Schwetz and A.~Zhou,
%``The fate of hints: updated global analysis of three-flavor neutrino oscillations,''
JHEP \textbf{09}, 178 (2020)
%doi:10.1007/JHEP09(2020)178
[arXiv:2007.14792 [hep-ph]], NuFIT 5.2 (2022).
%981 citations counted in INSPIRE as of 05 Dec 2023

\bibitem{Gonzalez_d5} 
%\cite{GonzalezGarcia:2012sz}
%\bibitem{GonzalezGarcia:2012sz} 
  M.~C.~Gonzalez-Garcia, M.~Maltoni, J.~Salvado and T.~Schwetz,
  %``Global fit to three neutrino mixing: critical look at present precision,''
  JHEP {\bf 1212}, 123 (2012)
  [arXiv:1209.3023v3 [hep-ph]], NuFIT 3.2 (2018).
  %%CITATION = ARXIV:1209.3023;%%
  %143 citations counted in INSPIRE as of 03 Jul 2013
%---------

\bibitem{Valle_d5}
%\cite{Tortola:2012te}
%\bibitem{Tortola:2012te} 
  D.~V.~Forero, M.~Tortola and J.~W.~F.~Valle,
  %``Global status of neutrino oscillation parameters after Neutrino-2012,''
  Phys.\ Rev.\ D {\bf 86}, 073012 (2012)
  [arXiv:1205.4018 [hep-ph]].
  %%CITATION = ARXIV:1205.4018;%%
  %251 citations counted in INSPIRE as of 25 Jul 2013

%******************************************************************

\bibitem{brp_d5} 
%\cite{Brahmachari:2012cq}
%\bibitem{Brahmachari:2012cq} 
  B.~Brahmachari and A.~Raychaudhuri,
  %``Perturbative generation of $theta_{13}$ from tribimaximal neutrino mixing,''
  Phys.\ Rev.\ D {\bf 86}, 051302 (2012)
  [arXiv:1204.5619 [hep-ph]];
  S.~Pramanick and A.~Raychaudhuri,
  %``Smallness of $\theta_{13}$ and the size of the solar mass splitting: Are they related?,''
  Phys.\ Rev.\ D {\bf 88},  093009 (2013)
  [arXiv:1308.1445 [hep-ph]].


\bibitem{pr_d5}   
  S.~Pramanick and A.~Raychaudhuri,
  %``Relating the small parameters of neutrino oscillations,''
  Phys.\ Lett.\ B {\bf 746}, 237 (2015)
  [arXiv:1411.0320 [hep-ph]];
  %``Relating small neutrino masses and mixing,''
  Int.\ J.\ Mod.\ Phys.\ A {\bf 30}, 1530036 (2015)
  [arXiv:1504.01555 [hep-ph]].
  %%CITATION = ARXIV:1504.01555;%%

\bibitem{seesaw_d5}
P. Minkowski,  Phys.\ Lett.\ B {\bf 67}, 421 (1977); 
M.~Gell-Mann, P.~Ramond and R.~Slansky,
in \textit{Supergravity}, p.~315, edited by F. van Nieuwenhuizen 
and D. Freedman, North Holland, Amsterdam, (1979); 
T.~Yanagida, Proc. of the \textit{Workshop on Unified Theory and 
the Baryon Number of the  Universe}, KEK, Japan, (1979); 
S.L. Glashow,  NATO Sci.\ Ser.\ B {\bf 59}, 687 (1980); 
R.N. Mohapatra and G.~Senjanovi{\'c},  Phys.\ Rev.\ D {\bf 23}, 165 (1981);
%\cite{Schechter:1981cv}
%\bibitem{Schechter:1981cv} 
  J.~Schechter and J.~W.~F.~Valle,
  %``Neutrino Decay and Spontaneous Violation of Lepton Number,''
  Phys.\ Rev.\ D {\bf 25}, 774 (1982);
%  doi:10.1103/PhysRevD.25.774
  %%CITATION = doi:10.1103/PhysRevD.25.774;%%
  %827 citations counted in INSPIRE as of 17 Apr 2019
%\cite{Schechter:1980gr}
%\bibitem{Schechter:1980gr} 
  J.~Schechter and J.~W.~F.~Valle,
  %``Neutrino Masses in SU(2) x U(1) Theories,''
  Phys.\ Rev.\ D {\bf 22}, 2227 (1980).
%  doi:10.1103/PhysRevD.22.2227
  %%CITATION = doi:10.1103/PhysRevD.22.2227;%%
  %2503 citations counted in INSPIRE as of 17 Apr 2019
  
  
%-----------
\bibitem{old_d5} 
%\bibitem{Vissani:1998xg} 
  F.~Vissani,
%``Large mixing, family structure, and dominant block in the neutrino mass matrix,''
JHEP {\bf 9811}, 025 (1998)
[hep-ph/9810435].
Models with  somewhat similar points of view as those
espoused here are 
%\bibitem{Akhmedov:1999ws} 
  E.~K.~Akhmedov,
  %``Small entries of neutrino mass matrices,''
Phys.\ Lett.\ B {\bf 467}, 95 (1999)
[hep-ph/9909217] and 
%\bibitem{Lindner:2007rs} 
  M.~Lindner and W.~Rodejohann,
  %``Large and almost maximal neutrino mixing within the type II see-saw mechanism,''
JHEP {\bf 0705}, 089 (2007)
[hep-ph/0703171].

\bibitem{othermodels_d5} For other recent work after the determination of
$\theta_{13}$ see 
%\cite{Antusch:2011ic}
%\bibitem{Antusch:2011ic} 
  S.~Antusch, S.~F.~King, C.~Luhn and M.~Spinrath,
  %``Trimaximal mixing with predicted $\theta_{13}$ from a new type of constrained sequential %dominance,''
  Nucl.\ Phys.\ B {\bf 856}, 328 (2012)
  [arXiv:1108.4278 [hep-ph]];
%\cite{Adhikary:2012mt}
%\bibitem{Adhikary:2012mt} 
  B.~Adhikary, A.~Ghosal and P.~Roy,
  %``$\theta_{13}$, $\mu\tau$ symmetry breaking and neutrino Yukawa textures,''
  Int.\ J.\ Mod.\ Phys.\ A {\bf 28},  1350118 (2013)
  arXiv:1210.5328 [hep-ph];
  %%CITATION = ARXIV:1210.5328;%%
  %3 citations counted in INSPIRE as of 01 Aug 2013
%\bibitem{Sierra:2013ypa} 
  D.~Aristizabal Sierra, I.~de Medeiros Varzielas and E.~Houet,
  %``An eigenvector based approach to neutrino mixing,''
  Phys.\ Rev.\ D {\bf 87}, 093009 (2013)
  [arXiv:1302.6499 [hep-ph]];
  %%CITATION = ARXIV:1302.6499;%%
  %1 citations counted in INSPIRE as of 01 Aug 2013
%\bibitem{Dutta:2013xla} 
  R.~Dutta, U.~Ch, A.~K.~Giri and N.~Sahu,
  %``Perturbative Bottom-up Approach for Neutrino Mass Matrix in Light of Large $\theta_{13}$ and Role of Lightest Neutrino Mass,''
  Int.\ J.\ Mod.\ Phys.\ A {\bf 29}, 1450113 (2014)
  arXiv:1303.3357 [hep-ph];
  %%CITATION = ARXIV:1303.3357;%%
  %2 citations counted in INSPIRE as of 01 Aug 2013
%\cite{Hall:2013yha}
%\bibitem{Hall:2013yha} 
  L.~J.~Hall and G.~G.~Ross,
  %``Stitching the Yukawa Quilt in the light of $\theta_{13}$,''
  JHEP {\bf 1311}, 091 (2013)
  arXiv:1303.6962 [hep-ph];
  %%CITATION = ARXIV:1303.6962;%%
  %3 citations counted in INSPIRE as of 01 Aug 2013
  T.~Araki,
  %``Is $\theta_{13}$^{PMNS} correlated with \theta_{23}^{PMNS} or not?,''
  PTEP {\bf 2013},  103B02 (2013)
  arXiv:1305.0248 [hep-ph];
  %%CITATION = ARXIV:1305.0248;%%
%\cite{Hernandez:2013dta}
%\bibitem{Hernandez:2013dta} 
  A.~E.~Carcamo Hernandez, I.~de Medeiros Varzielas, S.~G.~Kovalenko, H.~P\"{a}s and I.~Schmidt,
  %``Lepton masses and mixings in an $A_4$ multi-Higgs model with a radiative seesaw mechanism,''
  Phys.\ Rev.\ D {\bf 88},  076014 (2013)
  [arXiv:1307.6499 [hep-ph]];
  %%CITATION = ARXIV:1307.6499;%%
%\cite{Chen:2013wba}
%\bibitem{Chen:2013wba} 
  M.~-C.~Chen, J.~Huang, K.~T.~Mahanthappa and A.~M.~Wijangco,
  %``Large \theta_13 in a SUSY SU(5)xT' Model,''
  JHEP {\bf 1310}, 112 (2013)
  [arXiv:1307.7711] [hep-ph];
%\cite{Brahmachari:2014npa}
%\bibitem{Brahmachari:2014npa}
  B.~Brahmachari and P.~Roy,
  %``Testable constraint on near-tribimaximal neutrino mixing,''
  JHEP {\bf 1502}, 135 (2015)
  [arXiv:1407.5293 [hep-ph]];
  %%CITATION = ARXIV:1108.4278.%%
%\cite{BhupalDev:2012nm}
%\bibitem{BhupalDev:2012nm} 
  P.~S.~Bhupal Dev, B.~Dutta, R.~N.~Mohapatra and M.~Severson,
  %``$\theta_{13}$ and Proton Decay in a Minimal $SO(10) \times S_4$ model of Flavor,''
  Phys.\ Rev.\ D {\bf 86}, 035002 (2012)
%  doi:10.1103/PhysRevD.86.035002
  [arXiv:1202.4012 [hep-ph]].
  %%CITATION = doi:10.1103/PhysRevD.86.035002;%%
  %75 citations counted in INSPIRE as of 10 Apr 2019

\bibitem{LuhnKing_d5} For a review see, for example, S.~F.~King and C.~Luhn,
  %``Neutrino Mass and Mixing with Discrete Symmetry,''
  Rept.\ Prog.\ Phys.\  {\bf 76}, 056201 (2013)
  [arXiv:1301.1340 [hep-ph]].
  %%CITATION = ARXIV:1301.1340;%%
  %211 citations counted in INSPIRE as of 14 Aug 2015

\bibitem{ourS3_d5} 
  S.~Pramanick and A.~Raychaudhuri,
  %``Neutrino mass model with $S_3$ symmetry and seesaw interplay,''
  Phys.\ Rev.\ D {\bf 94}, no. 11, 115028 (2016)
%  doi:10.1103/PhysRevD.94.115028
  [arXiv:1609.06103 [hep-ph]],
  %%CITATION = doi:10.1103/PhysRevD.94.115028;%%
  %5 citations counted in INSPIRE as of 30 Oct 2017
%\cite{Pramanick:2017fdq}

\bibitem{newA4_d5} 
  S.~Pramanick,
  %``Ameliorating the popular lepton mixings with A4 symmetry: A seesaw model for realistic neutrino masses and mixing,''
  Phys.\ Rev.\ D {\bf 98}, no. 7, 075016 (2018)
%%  doi:10.1103/PhysRevD.98.075016
  [arXiv:1711.03510 [hep-ph]].
  %%CITATION = doi:10.1103/PhysRevD.98.075016;%%
%\cite{Pramanick:2015qga}

\bibitem{ourA4_d5} 
  S.~Pramanick and A.~Raychaudhuri,
  %``A4-based seesaw model for realistic neutrino masses and mixing,''
  Phys.\ Rev.\ D {\bf 93}, no. 3, 033007 (2016)
%  doi:10.1103/PhysRevD.93.033007
  [arXiv:1508.02330 [hep-ph]].
  %%CITATION = doi:10.1103/PhysRevD.93.033007;%%
  %8 citations counted in INSPIRE as of 30 Oct 2017

%\cite{Ma:2008ym}
\bibitem{Ma_rad_d5} 
  E.~Ma,
  %``Dark Scalar Doublets and Neutrino Tribimaximal Mixing from A(4) Symmetry,''
  Phys.\ Lett.\ B {\bf 671}, 366 (2009)
%  doi:10.1016/j.physletb.2008.12.038
  [arXiv:0808.1729 [hep-ph]].
  %%CITATION = doi:10.1016/j.physletb.2008.12.038;%%
  %50 citations counted in INSPIRE as of 06 Mar 2019
  
  
  


\bibitem{newMa_d5} 
%\cite{Ma:2011yi}
%\bibitem{Ma:2011yi} 
  E.~Ma and D.~Wegman,
  %``Nonzero theta(13) for neutrino mixing in the context of A(4) symmetry,''
Phys.\ Rev.\ Lett.\  {\bf 107}, 061803 (2011)
[arXiv:1106.4269 [hep-ph]];
%%CITATION = ARXIV:1106.4269;%%
%\cite{Gupta:2011ct}
%\bibitem{Gupta:2011ct} 
  S.~Gupta, A.~S.~Joshipura and K.~M.~Patel,
  %``Minimal extension of tri-bimaximal mixing and generalized Z_2 X Z_2 %symmetries,''
Phys.\ Rev.\ D {\bf 85}, 031903 (2012)
[arXiv:1112.6113 [hep-ph]];
%%CITATION = ARXIV:1112.6113;%%
%\cite{Dev:2012ns}
%\cite{Branco:2012vs}
%\bibitem{Branco:2012vs} 
  G.~C.~Branco, R.~G.~Felipe, F.~R.~Joaquim and H.~Serodio,
  %``Spontaneous leptonic CP violation and nonzero $\theta_{13}$,''
arXiv:1203.2646 [hep-ph];
%%CITATION = ARXIV:1203.2646;%%
  B.~Adhikary, B.~Brahmachari, A.~Ghosal, E.~Ma and M.~K.~Parida,
  %``A(4) symmetry and prediction of U(e3) in a modified Altarelli-Feruglio model,''
Phys.\ Lett.\ B {\bf 638}, 345 (2006)
[hep-ph/0603059];
%%CITATION = HEP-PH/0603059;%%
%\cite{Karmakar:2014dva}
%\bibitem{Karmakar:2014dva} 
  B.~Karmakar and A.~Sil,
  %``Nonzero $?_{13}$ and leptogenesis in a type-I seesaw model with $A_4$ symmetry,''
  Phys.\ Rev.\ D {\bf 91}, 013004 (2015)
  [arXiv:1407.5826 [hep-ph]];
  %%CITATION = ARXIV:1407.5826;%%
  %3 citations counted in INSPIRE as of 10 Aug 2015
%\cite{Ma:2015fpa}
%\bibitem{Ma:2015fpa} 
  E.~Ma,
  %``Neutrino Mixing: $A_4$ Variations,''
Phys.\ Lett.\ B {\bf 752}, 198 (2016)
%doi:10.1016/j.physletb.2015.11.049
[arXiv:1510.02501 [hep-ph]];
%\cite{He:2006dk}
%\bibitem{He:2006dk} 
  X.~G.~He, Y.~Y.~Keum and R.~R.~Volkas,
  %``A(4) flavor symmetry breaking scheme for understanding quark and neutrino mixing angles,''
  JHEP {\bf 0604}, 039 (2006)
%  doi:10.1088/1126-6708/2006/04/039
  [hep-ph/0601001].
  %%CITATION = doi:10.1088/1126-6708/2006/04/039;%%
  %266 citations counted in INSPIRE as of 09 Mar 2019
 

\bibitem{tanimoto_d5} 
%\cite{Kang:2015xfa}
%\bibitem{Kang:2015xfa} 
  S.~K.~Kang and M.~Tanimoto,
  %``Prediction of Leptonic CP Phase in $A_4$ symmetric model,''
Phys.\ Rev.\ D {\bf 91}, no. 7, 073010 (2015)
%doi:10.1103/PhysRevD.91.073010
[arXiv:1501.07428 [hep-ph]].

%%%%%%%%%%%%%%%%%%%%%%%%%%%%%%%%%%%%%%%%%%%%%%%%%%5
%\cite{Kang:2019sab}
\bibitem{scotogenic_Prof_Valle_d5} 
  Kang, O.~Popov, R.~Srivastava, J.~W.~F.~Valle and C.~A.~Vaquera-Araujo,
  %``Scotogenic dark matter stability from gauged matter parity,''
  arXiv:1902.05966 [hep-ph];
  %%CITATION = ARXIV:1902.05966;%%
  %1 citations counted in INSPIRE as of 17 Apr 2019
%\cite{Rojas:2018wym}
%\bibitem{Rojas:2018wym} 
  N.~Rojas, R.~Srivastava and J.~W.~F.~Valle,
  %``Simplest Scoto-Seesaw Mechanism,''
  Phys.\ Lett.\ B {\bf 789}, 132 (2019)
%  doi:10.1016/j.physletb.2018.12.014
  [arXiv:1807.11447 [hep-ph]];
  %%CITATION = doi:10.1016/j.physletb.2018.12.014;%%
  %5 citations counted in INSPIRE as of 17 Apr 2019
%\cite{Diaz:2016udz}
%\bibitem{Diaz:2016udz} 
  M.~A.~Díaz, N.~Rojas, S.~Urrutia-Quiroga and J.~W.~F.~Valle,
  %``Heavy Higgs Boson Production at Colliders in the Singlet-Triplet Scotogenic Dark Matter Model,''
  JHEP {\bf 1708}, 017 (2017)
%  doi:10.1007/JHEP08(2017)017
  [arXiv:1612.06569 [hep-ph]];
  %%CITATION = doi:10.1007/JHEP08(2017)017;%%
  %4 citations counted in INSPIRE as of 17 Apr 2019
%\cite{Merle:2016scw}
%\bibitem{Merle:2016scw} 
  A.~Merle, M.~Platscher, N.~Rojas, J.~W.~F.~Valle and A.~Vicente,
  %``Consistency of WIMP Dark Matter as radiative neutrino mass messenger,''
  JHEP {\bf 1607}, 013 (2016)
%  doi:10.1007/JHEP07(2016)013
  [arXiv:1603.05685 [hep-ph]];
  %%CITATION = doi:10.1007/JHEP07(2016)013;%%
  %20 citations counted in INSPIRE as of 17 Apr 2019
%\cite{Hirsch:2013ola}
%\bibitem{Hirsch:2013ola} 
  M.~Hirsch, R.~A.~Lineros, S.~Morisi, J.~Palacio, N.~Rojas and J.~W.~F.~Valle,
  %``WIMP dark matter as radiative neutrino mass messenger,''
  JHEP {\bf 1310}, 149 (2013)
%  doi:10.1007/JHEP10(2013)149
  [arXiv:1307.8134 [hep-ph]];
  %%CITATION = doi:10.1007/JHEP10(2013)149;%%
  %37 citations counted in INSPIRE as of 17 Apr 2019
%\cite{Bonilla:2016diq}
%\bibitem{Bonilla:2016diq} 
  C.~Bonilla, E.~Ma, E.~Peinado and J.~W.~F.~Valle,
  %``Two-loop Dirac neutrino mass and WIMP dark matter,''
  Phys.\ Lett.\ B {\bf 762}, 214 (2016)
%  doi:10.1016/j.physletb.2016.09.027
  [arXiv:1607.03931 [hep-ph]].
  %%CITATION = doi:10.1016/j.physletb.2016.09.027;%%
  %46 citations counted in INSPIRE as of 19 Jun 2019




%))))))))))))))))))))))))

%)))))))))))))))))))))

%\cite{Ma:2004br}
\bibitem{Ma_d5}
E.~Ma,
%``Polygonal derivation of the neutrino mass matrix,''
Fizika B \textbf{14}, 35-40 (2005)
[arXiv:hep-ph/0409288 [hep-ph]].
%27 citations counted in INSPIRE as of 06 Dec 2023

%\cite{Hagedorn:2006ir}
\bibitem{Lindner_d5}
C.~Hagedorn, M.~Lindner and F.~Plentinger,
%``The Discrete flavor symmetry D(5),''
Phys. Rev. D \textbf{74}, 025007 (2006)
%doi:10.1103/PhysRevD.74.025007
[arXiv:hep-ph/0604265 [hep-ph]].
%58 citations counted in INSPIRE as of 06 Dec 2023


%\cite{Cai:2017jrq}
\bibitem{radreview_d5} 
  Y.~Cai, J.~Herrero-García, M.~A.~Schmidt, A.~Vicente and R.~R.~Volkas,
  %``From the trees to the forest: a review of radiative neutrino mass models,''
  Front.\ in Phys.\  {\bf 5}, 63 (2017)
%%  doi:10.3389/fphy.2017.00063
  [arXiv:1706.08524 [hep-ph]];
  %%CITATION = doi:10.3389/fphy.2017.00063;%%
  %51 citations counted in INSPIRE as of 06 Mar 2019
%\cite{Klein:2019iws}
%\bibitem{Klein:2019iws} 
  C.~Klein, M.~Lindner and S.~Ohmer,
  %``Minimal Radiative Neutrino Masses,''
  arXiv:1901.03225 [hep-ph].
  %%CITATION = ARXIV:1901.03225;%%
  
  
%\cite{Pramanick:2019oxb}
\bibitem{radS3_d5}
S.~Pramanick,
%``Scotogenic S3 symmetric generation of realistic neutrino mixing,''
Phys. Rev. D \textbf{100}, no.3, 035009 (2019)
%doi:10.1103/PhysRevD.100.035009
[arXiv:1904.07558 [hep-ph]].
%19 citations counted in INSPIRE as of 06 Dec 2023


%\cite{Pramanick:2019qpg}
\bibitem{radA4_d5}
S.~Pramanick,
%``Radiative generation of realistic neutrino mixing with $A4$,''
Nucl. Phys. B \textbf{963}, 115282 (2021)
doi:10.1016/j.nuclphysb.2020.115282
[arXiv:1903.04208 [hep-ph]].
%14 citations counted in INSPIRE as of 06 Dec 2023

%***************************

%\cite{Ma:2006km}
\bibitem{Ma_loop_d5} 
  E.~Ma,
  %``Verifiable radiative seesaw mechanism of neutrino mass and dark matter,''
  Phys.\ Rev.\ D {\bf 73}, 077301 (2006)
%  doi:10.1103/PhysRevD.73.077301
  [hep-ph/0601225].
  %%CITATION = doi:10.1103/PhysRevD.73.077301;%%
  %881 citations counted in INSPIRE as of 06 Mar 2019

%\cite{Ishimori:2010au}
\bibitem{Ishimori_d5}
H.~Ishimori, T.~Kobayashi, H.~Ohki, Y.~Shimizu, H.~Okada and M.~Tanimoto,
%``Non-Abelian Discrete Symmetries in Particle Physics,''
Prog. Theor. Phys. Suppl. \textbf{183}, 1-163 (2010)
%doi:10.1143/PTPS.183.1
[arXiv:1003.3552 [hep-th]].
%956 citations counted in INSPIRE as of 10 Dec 2023



\end{thebibliography}
\end{document}